\documentclass[prd,aps,twocolumn,nofootinbib,superscriptaddress,eqsecnum,floatfix,preprintnumbers,amsmath,amssymb,longbibliography]{revtex4-2}

\usepackage{graphicx}
\usepackage{epsfig}
\usepackage[dvipsnames]{xcolor}
\usepackage[dvipsnames]{xcolor}
\usepackage[utf8]{inputenc}
\usepackage{stmaryrd}
\usepackage{mathrsfs}
\usepackage{mathalfa}
\usepackage{accents}
\usepackage{enumitem}
\usepackage[normalem]{ulem}

\usepackage[colorlinks=true,
            citecolor=green,
            linkcolor=red,
            filecolor=cyan,
            urlcolor=magenta,
            backref=false]{hyperref}

\newcommand{\ts}[1]{{#1}}

\newcommand{\la}{\langle}
\newcommand{\ra}{\rangle}

\newcommand{\be}{\begin{equation}}
\newcommand{\ee}{\end{equation}}
\newcommand{\bea}{\begin{eqnarray}}
\newcommand{\eea}{\end{eqnarray}}
\newcommand{\ba}{\begin{equation}\begin{aligned}}
\newcommand{\ea}{\end{aligned}\end{equation}}
\newcommand{\beg}{\begin{gather*}}
\newcommand{\eng}{\end{gather*}}
\newcommand{\hh}{,\hspace{0.5cm}}
\newcommand{\hhh}{,\hspace{0.2cm}}
\newcommand{\eq}[1]{\eqref{#1}}
\newcommand{\n}[1]{\label{#1}}
\newcommand{\ins}[1]{{\mbox{\tiny #1}}}

\newcommand{\CAL}{\mathcal}

\newcommand{\const}{\mbox{const}}

\begin{document}

\title{Regular black holes inspired by quasi-topological gravity}

\author{Valeri P. Frolov}
\email[]{vfrolov@ualberta.ca}
\affiliation{Theoretical Physics Institute, Department of Physics,
University of Alberta,\\
Edmonton, Alberta, T6G 2E1, Canada
}
\affiliation{
Center for Gravitational Physics and Quantum Information, Yukawa Institute for Theoretical Physics,
Kyoto University, 606-8502, Kyoto, Japan
}
\author{Alex Koek}
\email[]{koek@ualberta.ca}
\affiliation{Theoretical Physics Institute, Department of Physics,
University of Alberta,\\
Edmonton, Alberta, T6G 2E1, Canada
}
\author{Jose Pinedo Soto}
\email[]{pinedoso@ualberta.ca}
\affiliation{Theoretical Physics Institute, Department of Physics,
University of Alberta,\\
Edmonton, Alberta, T6G 2E1, Canada
}
\author{Andrei Zelnikov}%
\email[]{zelnikov@ualberta.ca}
\affiliation{Theoretical Physics Institute, Department of Physics,
University of Alberta,\\
Edmonton, Alberta, T6G 2E1, Canada
}


\begin{abstract}
Recently it was demonstrated that by adding to the Einstein-Hilbert action a series in powers of the curvature invariants with specially chosen coefficients one can obtain a theory of gravity which has spherically symmetric solutions describing regular black holes. Its reduced action depends on a function of one of the basic curvature invariants of the corresponding metric. In this paper we study a generalization of this model to the case when this function depends
on all the basic curvature invariants. We show that the metrics which are solutions of such a model possess a universal scaling property. We demonstrate that there exists a special class of such  models for which the ``master" equation for a basic curvature invariant is a linear second order ordinary differential equation. We specify a domain in the space of parameters of the model for which the corresponding solutions describe  regular static spherically symmetric black holes and study their properties.
\end{abstract}

{ \hfill    Alberta Thy 11-24}

\maketitle

\section{Introduction}

One of the assumptions of classical general relativity is the existence of standard ideal rulers and clocks which allow one to measure an interval between events and to introduce a metric. In quantum gravity the metric fluctuates. If $\hat{\ts{g}}$ is an operator describing a quantum gravitational field, one can write it in the form $\hat{\ts{g}}=\la \hat{\ts{g}}\ra+\delta \hat{\ts{g}}$. One can consider $\la \hat{\ts{g}}\ra$ as a quasi-classical background. If the fluctuations $\delta \hat{\ts{g}}$ are small, one can use $\la \hat{\ts{g}}\ra$ as an effective spacetime metric. This metric is obtained from some effective action, which in the low-curvature regime reduces to the Einstein-Hilbert action. In the high-curvature regime there exists a restriction on using such an effective action. The curvature describes the tidal force acting on an extended object by the gravitational field. In the very high-curvature regime this force  becomes so strong that no extended objects (such as rulers and clocks) can exist. One can consider this as evidence of the ultraviolet (UV) incompleteness of theory of gravity. As it is well known, in the framework of general relativity, such a situation occurs in the Big-Bang cosmology and in the interior of black holes. Classical solutions of the Einstein equations in both cases predict the existence of singularities, or spacetime domains where the curvature infinitely grows. One can interpret this as spacetime "terminating" its existence, at least in the form that we know. Another possibility is that a modification of Einstein's theory at high curvature results in the ``resolution" of these singularities. Instead of describing the formation of a singularity, such a theory would describe a regular geometry in its place, where the modified gravity still uses the notion of classical spacetime. Such a modification of gravity is important for the study of the final state of evaporating black holes.

There exist numerous publications in which different models of regular black holes are proposed and discussed.
Regular black hole solutions were studied in string theory \cite{Nicolini:2023hub}, in loop gravity \cite{Ashtekar2023}, in higher derivative and non-local gravity \cite{Buoninfante:2018xif,Buoninfante:2022ild,dePaulaNetto:2023cjw}, and in limiting curvature gravity \cite{Frolov:2021kcv,Frolov:2021afd}\footnote{The cited papers contain many useful references, but certainly do not cover all the relevant publications.}.
For simplicity, most of the publications consider spherically symmetric  black holes (see e.g.
 \cite{Ansoldi:2008jw, Frolov:2016pav,Carballo-Rubio:2018pmi} and references therein). Discussion of regular rotating black holes and corresponding references can be found in
\cite{Bambi_2013,ghosh2015,Simpson:2021dyo,Torres:2022twv,Franzin:2022wai,Lan:2023cvz,Frolov:2023jvi,Ghosh:2022gka,Franzin:2022wai,Carballo-Rubio:2022kad}.
The review \cite{Carballo-Rubio:2025fnc} gives a list of conceptual and observational implications  related to regular black holes and black hole mimickers.

Besides attempts to obtain regular black hole metrics as solutions of some modified gravity equations, there exists another direction of research for regular black holes. Namely, using different type of arguments, one proposes a special form of the metric and tries to describe general interesting new features of such ``solutions".  One of the first models of this type was proposed by Bardeen \cite{BARDEEN}. One of best known spherically symmetric metrics of this type was proposed later by Hayward \cite{Hayward:2005gi}. Among other interesting properties, it has the following one: It contains one length-scale parameter, $\ell$, which is used as a regulator for the curvature. Invariants that are quadratic in curvature $\CAL{R}^2$ calculated for this metric are uniformly bounded
\be \n{MARK}
|\CAL{R}^2|\le \dfrac{\CAL{B}}{\ell^4} ,
\ee
where $\CAL{B}$ is a dimensionless parameter depending on the choice of the invariant, but not depending on the mass parameter that enters the metric. In this sense, these metrics satisfy Markov's limiting curvature conjecture \cite{Markov:1982,Markov:1984ii,Brandenberger:1995hd}.

There exist several general features of static spherically symmetric regular black holes. They can be illustrated by considering a metric of the form
\be \n{fff}
ds^2=-N^2 f dt^2+\dfrac{dr^2}{f}+r^2 d\Omega^2\, ,
\ee
where $N=N(r)$ and $f=f(r)$.
For this metric, $f=(\nabla r)^2$.
Denote
\be
p=\dfrac{1-f}{r^2}\, .
\ee
As we shall see later, $p$ is one of the basic curvature invariants of the metric \eqref{fff}. For a regular black hole, the curvature is finite at $r=0$.
This means that near the origin, $f=1-p_0 r^2+...$, where $p_0=p(r=0)$ is a constant with dimensions of $[length]^{-2}$.
For positive $p_0$, the  metric near the center $r=0$ is close to the de Sitter one, and one often says that it has a de Sitter-like core  \cite{Frolov:1988vj,Dymnikova:1992ux,Dymnikova:2003vt}.
For such a metric, if $f$ decreases as $r$ increases from the origin, it can become negative.
If this happens, the point $r=r_->0$ where $f=0$ determines the inner horizon.
If we assume that the corresponding spacetime is asymptotically flat, there should exist another point $r=r_+>r_-$ where the outer horizon is located\footnote{In a more complicated case, equation $f=0$ may have more than 2 solutions or, when $r_{+}=r_{-}$, the horizon can be degenerate.}.
For a static black hole, the outer horizon coincides with the event horizon.

For a collapsing object, both branches of the apparent horizon are formed during the collapse. If the black hole completely evaporates, the inner- and outer-horizons eventually meet one another. A model of an evaporating black hole with a closed apparent horizon and corresponding conformal diagram was first discussed in the paper \cite{Frolov:1981mz}.

There exists a common problem in such models. The inner horizon has negative surface gravity, which results in a situation where mass inflation is possible \cite{Poisson:1989zz,Poisson:1990eh}.
One of the consequences of this effect is the following. If one considers a quantum field on the background of a regular evaporating black hole, then calculations show that when the
black hole evaporates completely, a huge burst of Hawking radiation is emitted at the moment of the black hole's disappearance \cite{Frolov:2016gwl,Frolov:2017rjz} (see also \cite{Carballo-Rubio:2022kad}). The energy release in this process can formally surpass the initial mass of the black hole. This means that in order to be realistic, these models should be modified and include the back reaction of the quantum radiation on the metric.

Hayward's model \cite{Hayward:2005gi} and other models similar to it have been widely discussed in connection with the information loss paradox (see e.g. \cite{Frolov:2014jva} and references therein). However, the corresponding metrics are not solutions of some (modified) gravity equations. Recently, there was interesting progress in this direction. Namely, a so-called quasi-topological theory of gravity was proposed, in which the Hayward metric and many other spherically symmetric regular black metrics can be obtained as solutions of a special class of gravity equations.

The quasi-topological gravity (QTG) model was first proposed in \cite{Oliva:2010eb, Myers:2010ru}, in which they added to the Einstein-Hilbert action not only the Gauss-Bonnet term to their action, but also a new term that is cubic in curvature.  This was achieved by adding all possible cubic curvature terms to the Lagrangian, and then cleverly choosing their couplings so that their corresponding equations of motion were second-order against an AdS background.  This was done intentionally to explore a wider class of AdS/CFT correspondences outside of Einstein gravity.  The model was called quasi-topological because this new cubic in curvature term gives trivial contributions in six dimensions, despite the fact that it does not correspond to a topological invariant.
This class of theories was further generalized \cite{Oliva:2011xu,Oliva:2012zs,Oliva:2010zd} to arbitrarily higher orders in curvature.

Expanding on this, it was discovered by Bueno and Cano \cite{Bueno:2016cu} that there is only one unique theory of cubic curvature gravity that can propagate a transverse, massless, ghost-free graviton on a maximally symmetric background, which they called “Einstein Cubic Gravity”.  With collaborators, they extended this construction up to quartic order in curvature \cite{Bueno:2017ho}.

From this came the idea of generalized quasi-topological gravity (GQTG) \cite{Hennigar:2017ego}, which started off as the most general cubic theory for vacuum spherically symmetric spacetimes that produces the following remarkable properties: 1) It has a smooth limit in which it recovers Einstein gravity with a cosmological constant.  2) The theory is defined by a single metric function.  3) This metric function obeys a second-order differential equation in the vacuum, or with a suitable choice of matter.  4) The linearized theory propagates the same degrees of freedom as Einstein’s equations on a maximally symmetric background.  This theory was also generalized to be applicable at all orders of curvature \cite{Bueno:2020gq}.  There is strong evidence that all higher-order curvature gravities can be characterized as GQTG theories via field redefinitions \cite{Bueno:2019ltp}.  These GQTGs were further classified in \cite{Bueno:2022res,Moreno:2023rfl}, and it was shown that within these GQTGs there is only one unique QTG at each order of curvature such that the equation for the metric function is algebraic.

More recently, it was discovered that taking an infinite number of quasi-topological terms in the action, namely one for each order of curvature, produces a metric that is singularity-free at the origin \cite{Bueno:2024dgm}, but this is only possible if an infinite series of these terms is included.  For specific choices within this model, one can reproduce the Hayward metric for a number of dimensions greater than 4, without the inclusion of matter. In the presence of pressureless  matter in the form of collapsing thin shells it was recently shown \cite{Bueno:2024zsx,Bueno:2024eig} that these theories lead generically to the formation of regular black holes with the interior dynamics consisting of shell bounces and white hole explosions into a new universe.

These authors used a static spherically symmetric reduction of the action, which for the metric \eqref{fff} has the form
\be \n{TGS}
S=-B\int dt dr \dfrac{dN}{dr} r^{D-1} h(p)\, ,
\ee
where $h(p)$ is a scalar function of the curvature invariant $p$, and $B$ is a constant which depends on the gravitational coupling constant and the number of spacetime dimensions.

This form of the reduced action is a starting point of our model. We first demonstrate that it can be presented in the form of a dilaton gravity action in two dimensions, with $\ln r$ playing the role of the dilaton field. After this, we propose a generalization of such a dilaton gravity model by including other basic curvature invariants in the argument of the function $h$.

The paper is organized as follows. In Sec.\ref{sec2} we
remind the reader  the method of spherical reduction of the gravity action on the example of Einstein gravity and define the basis of curvature invariants.  In Sec.\ref{sec3} we propose a dilaton gravity model where $h$ is a non-linear function of basic curvature invariants. We demonstrate that metrics of the form \eqref{fff} which are solutions of the corresponding equations have a remarkable scaling property. We use this property to define a new scaling-invariant coordinate $y=y(r)$ in which the equations are greatly simplified.  In Sec.\ref{sec4} we consider a subclass of the proposed non-linear models in which the governing equation becomes an linear second-order ordinary differential equation (ODE), and we obtain explicit analytic solutions of this ``master" equation. We demonstrate that
for the special form of the
function $h=h(p)$,
the model is equivalent to quasi-topological gravity (QTG) \cite{Myers:2010ru,Hennigar:2017ego,Bueno:2019ltp,Bueno:2022res,Moreno:2023rfl,Bueno:2024dgm,DiFilippo:2024mwm}. We show that for a wide range of the parameters which enter the effective action proposed in this paper, the solutions are regular and their spacetime curvature invariants satisfy the inequality \eqref{MARK}. We  showed that obtained solutions include the Hayward metric \cite{Hayward:2005gi} which is reproduced for a specific choice of parameters of the theory. In Sec.\ref{sec5} we study regular black hole solutions of the propose model,as well as their properties.  The discussion of the obtained results is contained in Sec.\ref{discussion}. In the appendices we collect useful mathematical relations that are used in our computations.

In this paper we use the sign convention adopted in the book \cite{MTW}.


\section{Spherical reduction}\label{sec2}

\subsection{$S$-reduction}

In this paper we focus on spherically symmetric solutions of the modified gravity equations. For a covariant theory of gravity, such solutions can be obtained from a corresponding reduced action.

As a simple illustration of the symmetry reduction approach, let us describe how this method works in standard general relativity. Consider the Einstein-Hilbert action in $D$-dimensional spacetime
\be\n{EHA}
S[\,g\,]=\frac{1}{2\varkappa}\,\int d^D x \,\sqrt{-g}\, R\, .
\ee
Here $\varkappa$ is the $D$-dimensional gravitational coupling constant.
Suppose one searches for a spherically symmetric solution of the vacuum  Einstein equations\footnote{The material presented in this section is well known. We collect it here mainly in order to explain definitions and notations used later in the paper. This also allows us to describe basic curvature invariants of spherically symmetric spacetime which play an important role in the paper.}
\be \n{EINEQ}
G_{\mu\nu}\equiv R_{\mu\nu}-\dfrac{1}{2}g_{\mu\nu}R=0\, .
\ee
The metric of a spherically symmetric $D$-dimensional spacetime can be written in the form
\ba\n{metric}
ds^2=\tilde{g}_{ij}dx^i dx^j+\rho^{2}d\Omega^2\hhh  i,j=0,1\, .
\ea
Here the two-dimensional metric $\tilde{g}_{ij}$ and the variable $\rho$ are functions of $x^i$ coordinates, and
\ba\n{sphere}
d\Omega^2=\hat{g}_{ab}(y)dy^a dy^b\hhh  a,b=2,\dots,(D-1)
\ea
is the metric of a unit $(D-2)$-dimensional sphere $S^{D-2}$. Its area is
\be
\Omega_{D-2}=\dfrac{2\pi^{(D-1)/2}}{\Gamma[(D-1)/2]}\, .
\ee

In order to study spherically symmetric spacetimes, it is sufficient to substitute the
metric ansatz \eqref{metric}-\eqref{sphere} into the gravity equations \eqref{EINEQ}. This gives the following equations $G_i^j=0$ and $G_a^b=\delta_a^b V=0$, where
\ba\n{2d}
G_i^j&=-(D-2)\Big\{\frac{\rho_{:i}^{:j}}{\rho}\\
&+\delta_{i}^{j}\Big[-\frac{\rho_{:k}^{:k}}{\rho}+\frac{D-3}{2\rho^2}\big(1-\rho^{:k}\rho_{:k}\big)\Big]\Big\} =0 ,
\ea
\ba\n{V}
V=&-\frac{1}{2} \CAL{R}+(D-3)\frac{\rho_{:k}^{:k}}{\rho}\\
&-\frac{(D-3)(D-4)}{2\rho^2}\big(1-\rho^{:k}\rho_{:k}\big)=0\, .
\ea
Here $\CAL{R}$ is the scalar curvature of the two-dimensional metric $\tilde{g}_{ij}$ and a colon denotes the two-dimensional covariant derivative with respect to this metric.

The equation \eqref{V}, in fact, is not an independent equation but a consequence of \eqref{2d}, due to the relation
\ba
G_i^k{}_{:k}+(D-2)\,G_i^k\frac{\rho_{:k}}{\rho}=(D-2)\frac{\rho_{:i}}{\rho} V  .
\ea
This property is due to the Bianchi identity leading to the $D$-dimensional `conservation' law $G_{\alpha}^{\epsilon}{}_{;\epsilon}=0$.

However there is an alternative
way to go about this. Namely, one can directly substitute the ansatz \eqref{metric} into the
gravity action \eqref{EHA}. After integration over the angular variables, one obtains
the following reduced action
\ba
S_{S}=&\dfrac{\Omega_{D-2}}{2\varkappa}\int d^2 x \,\sqrt{-\tilde{g}}\, [\CAL{R}\rho^{D-2} \\ &+(D-2)(D-3)\rho^{D-4}(\rho^{:i}\rho_{:i}+1)]\, .
\ea

The reduced functional $S_{S}$ depends on
the two-dimensional metric $\tilde{g}_{ij}$ and a scalar function $\rho$,
both of which depend only on the $x^i$ coordinates, so that the original four-dimensional problem is reduced to a two-dimensional one.

Let us notice that in the general case, a substitution of some chosen ansatz for the metric into the action does not guarantee that after the variation of the reduced action,
the obtained equations will correctly reproduce the reduced Einstein
equations. The reason is that the variety of possible variations $\delta g_{\mu\nu}(x^{\alpha})$
of the original action is much larger than the variety of allowed variations of
the reduced action variables $\delta g_{ij}(x^k)$ and $\delta \rho(x^k)$ restricted by the symmetry constraints. However, one can check that for the  case of spherical symmetry, the variation of the reduced action correctly reproduces a complete set of the reduced gravity equations (see e.g. \cite{FZ}). We call the procedure described in this subsection a spherical reduction, and use a subscript $S$ to indicate that a corresponding quantity is obtained by means of this reduction.

\subsection{$SS$-reduction}

In the vacuum, the spherically reduced equations \eqref{2d} and \eqref{V} imply that the quantity
\be
\xi^{\mu}=\delta^{\mu}_{i} e^{ij}\rho_{:j}
\ee
is a Killing vector of the metric \eqref{metric} (Birkhoff’s theorem).
Here $e^{ij}$ is a two-dimensional Levi-Civita tensor in the space with In the general case with the presence of a spherical distribution of matter, the corresponding metric can be non-static. However, if one still wants to search for static spherically symmetric solutions of Einstein's equations, one can proceed with a further reduction. Namely, one assumes that the metric is both spherically symmetric and static, and so it admits the symmetry group $SO(D-1)\times R^1$.
A general form of the static spherically symmetric metric is
\ba\label{ds}
ds^2=-N^2 f dt^2+f^{-1}dr^2+\rho^2 d\Omega^2,
\ea
where $d\Omega^2$ is once again the line element on the $(D-2)$-dimensional unit sphere.
This metric contains three arbitrary functions of one variable, $N(r)$, $f(r)$, and $\rho(r)$.

Let us discuss the properties of static spherically symmetric metrics in more detail.
For such metrics, there exist four independent curvature invariants, which we denote by $p,q,u,v$, such that the $D$-dimensional curvature tensor components have the form
\ba\n{RRRR}
R_{jl}{}^{ik}&=v\,\delta_{jl}^{ik} ,\\
R_{ta}{}^{tb}&=u\,\delta_{a}^{b} ,\\
R_{ra}{}^{rb}&=q\,\delta_{a}^{b} ,\\
R_{ab}{}^{cd}&=p\,\delta_{ab}^{cd} ,
\ea
where $ i,j=0,1$ and $~a,b=2,\dots,(D-1)$. For the metric \eqref{ds}, these quantities have the following form
\ba\label{pquv0}
p&=\frac{1-(\rho')^2 f}{\rho^2} ,\\
q&=-\Big(\frac{f'}{2}\frac{\rho'}{\rho} +f \frac{\rho''}{\rho}\Big) ,\\
u&=-\Big(\frac{f'}{2}\frac{\rho'}{\rho} +f \frac{\rho'}{\rho}\frac{N'}{N}\Big) ,\\
v&=-\frac{1}{2}\Big(f''+3f'\frac{N'}{N}+2f\frac{N''}{N}\Big) .
\ea
Here and later on, we denote by a prime the derivative with respect to $r$.
We call functions $p$, $q$, $u$ and $v$ basic curvature invariants.
Other local scalar invariants constructed from the curvature can be expressed as a function of these basic curvature invariants\footnote{
As has been proved by Narlikar and Karmarkar \cite{NARLIKAR},  for  spherically symmetric metrics
which depend on both space and time coordinates, a set of four invariants is sufficient for the construction of all other algebraic invariants of the curvature tensor.
The metric \eqref{ds} belongs to a wider class of
so-called warped product metrics. A discussion of the complete
set of curvature invariants which is sufficient for construction
of algebraic curvature invariants and further references can be
found in \cite{CURV_INV_1998}.}.
For example, the Ricci scalar has the form
\ba\label{R}
R=&\big[v+(D-2)q\big]+\big[v+(D-2)u\big]\\&+(D-2)\big[(D-3)p+q+u\big] .
\ea

If one substitutes the expression \eqref{R} for the curvature into the Einstein-Hilbert action, then after integration by parts one obtains the corresponding reduced action

\ba\label{SE}
S_{SS}&=B\int dt dr\,N\rho^{D-2} [(D-3)p+2q]\\
&=B\int dt dr\, \frac{N}{\rho'}\frac{\partial}{\partial r}\big[\rho^{D-3}\big(1-(\rho')^2 f\big)\big]\\
&=B\int dt dr\, \frac{N}{\rho'}\frac{\partial}{\partial r}\big[\rho^{D-1}p\big] .
\ea
Here the constant $B$ is given by
\be\label{defB}
B=\frac{(D-2)\Omega_{D-2}}{2\varkappa} .
\ee
After integration by parts, we obtain
\be\n{LLL}
\begin{split}
S_{SS}&=B\int dt dr \CAL{L}\, ,\\
\CAL{L}&=-\CAL{W} p\hh
\CAL{W}=\rho^{D-1}\Big(\frac{N}{\rho'}\Big)'\, .
\end{split}
\ee

By varying this action over $f(r)$, one obtains the following  equation
\be \n{EQN}
\Big[\frac{N}{\rho'}\Big]'=0 .
\ee
It has a trivial solution
\ba\n{solN}
N=\alpha \rho'
\ea
where $\alpha$ is an arbitrary constant, which can be put equal to 1 by properly normalizing clocks at infinity.

Variation over the lapse function $N(r)$ gives the equation for $f$
\ba\n{EQf}
\big[\rho^{D-1}p\big]'=\big[\rho^{D-3}\big(1-(\rho')^2 f\big)\big]'=0 ,
\ea
with the solution
\ba\label{solf}
f=\frac{1}{(\rho')^2}\Big(1-\frac{\mu}{\rho^{D-3}}\Big).
\ea
This is the function which enters the Schwarzschild-Tangerlini metric
for a $D$-dimensional black hole. The constant of integration $\mu$ is  related to the scale parameter of the solution: the radius of the black hole $r_\ins{g}=\mu^{1/(D-3)}$. In $D$ dimensions, $\mu$ is proportional to the black hole mass $M$
\ba\label{muM}
\mu=\frac{2\varkappa M}{(D-2)\Omega_{D-2}}=\frac{1}{B}M .
\ea
In four dimensions it reproduces the familiar expression $r_\ins{g}=\mu=\frac{\varkappa M}{4\pi}=2G M$, where $G$ is the 4-dimensional gravitational constant. Note that we use units where the speed of light $c=1$.

The third equation is obtained  by variation of the action over the function $\rho(r)$. However, from the form of the action \eqref{SE}
\ba\label{SE1}
S_{SS}&=B\int dt dr\, H(N,\rho')\frac{\partial}{\partial r}\big[Q(f,\rho,\rho')\big]
\ea
it is clear that the equations
\eqref{EQN} and \eqref{EQf}
have the form $H'=0$ and $Q'=0$, while the third equation
takes the form
\ba
-\Big(\frac{\partial H}{\partial\rho'} Q'\Big)'+\Big(\frac{\partial Q}{\partial\rho'} H'\Big)' - \frac{\partial Q}{\partial\rho}H'=0 .
\ea
It happens to be a linear combination of the first two equations and their derivatives, so it is automatically satisfied and does not restrict any other parameters of the solution.

Therefore, without loss of generality we can choose the gauge
\ba
\rho=r .
\ea
In this choice of the coordinates we get a simple form for the solutions to \eq{solN}-\eq{solf}
\ba
N=1
\ea
and
\ba
f=1-\frac{\mu}{r^{D-3}} .
\ea
The corresponding solution is the Schwarzschild-Tangherlini metric.

We call the procedure described in this subsection a static spherical reduction, and use a subscript $SS$ to indicate that a corresponding quantity is obtained by means of this reduction.

\subsection{$SSR$-reduction}

The metric \eqref{ds} still has a freedom of coordinate transformation $r\to \tilde{r}(r)$ which preserves its form.
One can fix this freedom by choosing $\rho=r$. In such radial coordinates, the metric takes the standard form
\be \n{SSR1}
ds^2=-N^2 f dt^2+f^{-1}dr^2+r^2 d\Omega^2\, ,
\ee
and it contains two functions of one variable, $f(r)$ and $N(r)$. The Einstein equations in this gauge are
\begin{eqnarray}
&&N'=0\label{eq1} ,\\
&&(D-3)\big[1- f\big]- r f'=0 . \label{eq2}
\end{eqnarray}
The corresponding reduced action is obtained
by substituting the metric \eqref{SSR1} into the Einstein-Hilbert action \eqref{SE}.
Then, after integrating by parts and omitting surface integrals which do not contribute to the equations of motion, we get
\ba\label{SSR}
S_{SSR}&=-B\int dt dr\, N' r^{D-1} p\\
&=-B\int dt dr\, N' \big[r^{D-3} (1-f)\big] .\\
\ea
It depends on two functions of one variable, $f(r)$ and $N(r)$. The variation of $S_{SSR}$ over $f(r)$ gives \eq{eq1} while variation over $N(r)$ reproduces \eq{eq2}.
In fact, these two equations are equivalent to the set of equations \eq{EQN} and \eq{EQf} written in the gauge $\rho=r$.

We call the procedure described in this subsection a radial static spherical reduction, and use a subscript $SSR$ to indicate that a corresponding quantity is obtained by means of this reduction.

In the above examples, we demonstrated that spherically symmetric solutions of Einstein's equations can be obtained from the spherically reduced Einstein-Hilbert action. In fact, this is not a special property of the Einstein equations, but a generic property of Lagrangian field theories. There exists a so called Palais' Principle of Symmetric Criticality (PSC) that is applicable to the Lie symmetry reduction of Lagrangian field theories \cite{Palais:1979rca}. This principle states
that ``for any group-invariant Lagrangian the equations obtained by restriction of Euler-Lagrange equations to group-invariant fields are equivalent to the Euler-Lagrange equations of a canonically defined, symmetry-reduced Lagrangian" \cite{Fels:2001rv}. The conditions of validity of this principle for local gravitational theories built from a metric were discussed in \cite{Fels:2001rv,Anderson:1999cn}. In particular, it was demonstrated that for the spherical reduction of Einstein gravity, PSC is valid\footnote{See Section 6.3 (page 28) in \cite{ Fels:2001rv} and Example 6.5  in \cite{Anderson:1999cn}. For more information see also the recent paper \cite{Frausto:2024egp}.
Note that these results were explicitly proven for four-dimensional cases. However, the approach was developed for arbitrary dimensions and similar results concerning the validity of PSC are expected to remain valid in higher dimensions.
}.

Let us emphasize: the described ``from top to bottom" approach, where one starts with a covariant action and reduces it using the symmetry property of the theory, is a well-defined procedure. An interesting question is, under which conditions does the inverse ``from bottom to top" procedure has a solution? Namely, suppose one has an effective action which produces the equations for a chosen ansatz of the metric. Does there exist a covariant  action with a proper reduction that reproduces these equations?
We will address this question in the last subsection  \ref{SS_3C} of the next section.


\section{2D dilaton gravity model with a non-linear in curvature action }\label{sec3}

\subsection{Action}

In what follows, we consider static spherically reduced metrics \eqref{ds} and modify the reduced Einstein-Hilbert action \eqref{LLL}
\ba\n{SLL}
S&=B\int dt dr \sqrt{-g} L\\
&=B\int dt dr \mathcal{L}
\ea
as follows. We write the Lagrangian density $\CAL{L}$ in the form
\be \n{hhh}
\CAL{L}=-\CAL{W} h \, ,
\ee
where $\CAL{W}$ is\footnote{
There is another equivalent form of this expression
\ba\nonumber
\CAL{W}=\sqrt{-g}\,\frac{(q-u)\rho^2}{(\nabla\rho)^2}=\sqrt{-g}\,\frac{(q-u)}{(\nabla\varphi)^2}\, ,
\ea
where $\varphi=\ln \rho$.
}
\ba\n{WWWW}
\CAL{W}=\sqrt{-g}\,\frac{(q-u)\rho^2}{1-\rho^2 p}\, ,
\ea
and $h$ is is a function of the curvature invariants
\be \n{hpqvu}
h=h(p,q,u,v)\, .
\ee
Using the expressions for the curvature invariants \eqref{pquv0}, one can check that $\CAL{W}$ coincides with the expression given in \eqref{LLL}.

Thus, we shall consider equations of motion which are obtained by variation of the action
\be\n{ACT}
\begin{split}
S=&-B\int dt dr \CAL{W}h\\
=&-B\int dt dr\sqrt{-g}\frac{(q-u)\rho^2}{1-\rho^2 p}h\, ,
\end{split}
\ee
where $h$ is defined by \eqref{hpqvu}.
For $h=p$, this action reduces to the Einstein-Hilbert action.
For $h=h(p)$, it reproduces
the  action proposed in quasi-topological models \cite{Bueno:2022res}.
In this sense, the  action \eqref{ACT} is a natural generalization of these models.
The action \eqref{ACT} can be interpreted as an action for a two-dimensional non-linear in curvature dilaton gravity, where the dilaton field $\varphi$ is given by $\varphi=\ln\rho$.
\footnote{
For a general discussion of two-dimensional dilaton gravity see e.g. \cite{GRUMILLER2002327}. Regular black hole solutions in dilaton gravity were discussed in \cite{LOUISMARTINEZ1994193,Barenboim:2024dko}. Let us emphasize that only a special type of two-dimensional dilaton gravity can be obtained via the spherical reduction of a covariant four-dimensional modified gravity theory. It would be very interesting to find what the conditions that permit this reduction are. In some sense, this is the inverse problem of the PSC problem discussed in the previous section.
}

Let us emphasize that we do not require that this two-dimensional action is obtained by a spherical reduction of a covariant four-dimensional action. In this paper our main focus is connected with obtaining regular black hole metrics, and we use action \eqref{ACT} as a ``generator" for such solutions.

Variation of the action \eqref{ACT} over $f$ leads to the equation
\ba\n{Wf}
\frac{\partial h}{\partial f}\CAL{W}-\Big[\frac{\partial h}{\partial f'}\CAL{W}\Big]'
+\Big[\frac{\partial h}{\partial f''}\CAL{W}\Big]''=0\, .
\ea
We assume that $h$ only depends on $f$ and its derivatives up to second order, which is always the case if $h$ is a function of basic curvature invariants.
Similarly, the variation over $N$ gives the equation for $f$
\ba\n{WN}
&\frac{1}{\rho'}(\rho^{D-1} h )'\\
&-\frac{\partial h}{\partial N}\CAL{W}+\Big[\frac{\partial h}{\partial N'} \CAL{W}\Big]'
-\Big[\frac{\partial h}{\partial N''}\CAL{W}\Big]''=0 .
\ea
Equations \eqref{Wf} and \eqref{WN} should be ``equipped" with appropriate boundary conditions. In the general case, these equations are nonlinear and the proper choice of such conditions might be highly non-trivial. Instead of worrying about this, we proceed as follows. Let us note that \eqref{Wf} always has a simple solution $\CAL{W}=0$ for which
$N/\rho'=\const$.
For this solution, the second line in the equation vanishes and it greatly simplifies and reduces to
\ba\n{HHHH0}
(\rho^{D-1} h)'=0 .
\ea
Its solution is
\ba\n{hhhh}
h=\frac{\tilde{\mu}}{\rho^{D-1}}\, ,
\ea
where $\tilde{\mu}$ is an integration constant. In what follows we focus on this class of solutions. The constant $\tilde{\mu}$ defines asymptotic behavior of the solution at large distance and, hence, is proportional to the mass parameter $\mu$ of the solution $\tilde{\mu}=Z\mu$. The coefficient of proportionality $Z$ depends on the particular choice of the theory parameters. Recall that in $D$ dimensions the mass of the black hole $M$ and the mass parameter $\mu$ are related according to (\ref{muM}).

The third equation coming from the variation over the `dilaton' $\rho$ is satisfied automatically as a consequence of the first two equations.

Now we can use the gauge $\rho=r$, and fix normalization of the lapse function $N=1$ at infinity. Then we get the solution
\be \label{N}
N=1\hh
h=\frac{\tilde{\mu}}{r^{D-1}}\, .
\ee

\subsection{Scaling property}\n{SS_3B}

The basic curvature invariants $p$, $q$, $u$, and $v$ have dimensions of $1/[Length]^2$. This follows both from their definition \eqref{RRRR}, and the explicit form \eqref{pquv0}. The action \eqref{ACT} for $h=p$ reproduces the reduced Einstein-Hilbert action. Hence, $h$ also has dimensions of $1/[Length]^2$. In what follows we consider $h$ in the case where it is a non-linear function of the basic curvature invariants. Suppose we choose $h=p+\beta p^2$. In order to have proper dimensionality, the coefficient $\beta$ should have dimensions of $[Length]^2$. This means that in the general non-linear case, a theory should possess some fundamental length, which we denote in our case by $\ell$ \footnote{
If the theory has several length parameters, we choose one of them as $\ell$, and the ratio of the other length parameters and $\ell$ enter as dimensionless parameters of the model.
}.
In quantum gravity, there exists a natural length parameter-- the Planck length $\ell_\ins{Pl}$. In this paper we do not assume that $\ell=\ell_\ins{Pl}$ and
keep $\ell$ arbitrary. In the non-linear case, we consider $\ell$ as some fundamental length which defines a scale at which non-linearity becomes important.

From \eqref{pquv0} written in the gauge $\rho=r$ with $N=1$, we get the following differential relations between basic curvature invariants
\ba\label{invars1}
q&=u=p+\frac{1}{2}r\partial_r p\, ,\\
v&=p+2r\partial_r p+\frac{1}{2}r^2\partial^2_r p\, .
\ea
Substituting these expressions into \eqref{hhhh} and using \eqref{N}, one gets
\be \n{HHHH}
h\equiv h(r,p,p',p'')=\dfrac{\tilde{\mu}}{r^{D-1}}\, .
\ee
This is an ODE for $p$. Its solution is a function of $r$ and two parameters, $\tilde{\mu}$ and  $\ell$,
\be
p=p(r,\tilde{\mu},\ell)\, .
\ee
Since the dimension of $p$ is $[Length]^{-2}$, the Buckingham $\pi$ theorem implies that it can be written in the form
\be \n{PPPP}
p=\ell^{-2} P(r/\ell,\tilde{\mu}/\ell^{D-3})\, .
\ee

Let us demonstrate that one can find a new coordinate $y$, such that the basic invariant $p$ can be written as function of one variable $p=p(y)$. This property follows from a special form of the differential operators in \eqref{invars1}. Let us denote
\ba\n{yyy}
y=\frac{r^{D-1}}{\ell^2  \tilde{\mu}}\, .
\ea
Then \eqref{invars1} can be written as follows
\ba\label{invars}
q&=u=p+\frac{(D-1)}{2}y\partial_y p\, ,\\
v&=p+\frac{(D-1)(D+2)}{2}y\partial_y p
+\frac{(D-1)^2}{2} y^2\partial^2_y p\, .
\ea
Multiplying both sides of \eqref{HHHH} by $\ell^2$, one obtains the following equation
\be \n{EEQQ}
\hat{h}(y,\hat{p},d\hat{p}/dy,d^2\hat{p}/dy^2)=\dfrac{1}{y}\, ,
\ee
where
\be
\hat{p}=\ell^2 p \hhh
\hat{h}=\ell^2 h \, .
\ee
This means that, when written in terms of a new dimensionless coordinate $y$, the equation \eqref{EEQQ} does not contain parameters $\ell$ and $\tilde{\mu}$, and in this sense it is universal. We refer to this property of our model as the scaling property. We also call \eqref{EEQQ} a master equation.

\subsection{Covariant action}\n{SS_3C}

Is it possible to reconstruct a covariant $D$-dimensional action which after spherical reduction reproduces the $2D$ dilaton action (\ref{ACT})? The answer is yes, provided the function $h$ (\ref{hpqvu})  depends only on the invariant $p$. To prove this, consider the action (\ref{ACT}) in the $\rho=r$ gauge, integrate it by parts, and omit surface terms because they do not affect the equations of motion. Then the reduced action takes the form
\be\n{ACT1}
S=B\int dt dr\, N\big[r^{D-1}h( p)\big]'  .
\ee
The function $N$ can be expressed in terms of $D$-dimensional metric density $\sqrt{-g}=N r^{D-2}\sqrt{\hat{g}}$, where $\hat{g}=\det(\hat{g}_{ab})$ corresponds to the metric (\ref{sphere}) on the $(D-2)$-dimensional unit sphere.

Firstly let us consider  the following combination
\ba
 \frac{1}{r^{D-2}}\frac{\partial}{\partial r}\big[r^{D-1}h( p)\big]&=(D-1)h + r\frac{\partial}{\partial r}h( p)\\
 &=(D-1)h +r p'\frac{\partial}{\partial p}h( p)  .
 \ea
Using (\ref{pquv0}) in the gauge $\rho=r$ we obtain the identity
\be
rp'=2(q-p).
\ee
Then one can see that
\ba\nonumber
\sqrt{-g}\Big[(D-1)h +2(q-p)\frac{\partial h( p)}{\partial p}\Big]
=\sqrt{\hat{g}}N\big(r^{D-1}h( p)\big)' .
\ea
The left-hand side of this expression has a completely $D$-dimensional covariant form, and can be written explicitly in terms of curvature invariants $R$, $C_{\alpha\beta\mu\nu} C^{\alpha\beta\mu\nu}$, $R_{\mu}^{\nu} R_{\nu}^{\mu}$, and $R_{\mu}^{\nu} R_{\nu}^{\sigma}R_{\sigma}^{\mu}$ using formulas in appendix \ref{AppInv}.

It means that for any given function $h(p)$, the covariant action
\ba\n{EHAcov}
S[\,g\,]=&\frac{1}{2\varkappa}\,\int d^D x \,\sqrt{-g}\Big[(D-1)h+2(q-p)\frac{\partial}{\partial p}h \Big] ,
\ea
after the spherical reduction reproduces our ansatz (\ref{ACT}) for the $2D$ dilaton action.

In more general cases, e.g., with $h(p,q)$, the problem of reconstructing a covariant action is more cumbersome,
and should be subject to further study.


\section{Master equation and its solutions}\label{sec4}

\subsection{Choice of function $h$}

In the general case, the equation \eqref{EEQQ} is a non-linear second order ODE. To simplify it, we shall make the following two assumptions: i) $h$ is a function $h(\psi)$ of the following linear combination of basic curvature invariants
    $\psi=ap+bq+cv$;
 ii) This function has the following form
    \be
h(\psi)=\frac{\psi}{1-\ell^2 \psi}\, .
    \ee

The second assumption allows one to reduce \eqref{EEQQ} to the following equation
\be \n{ppss}
\ell^2\psi=\frac{1}{1+y}\, ,
\ee
while the first assumption implies that \eqref{ppss} has the following form, which is linear in the basic curvature invariants
\ba\n{FEQ}
ap+bq+cv=\frac{1}{\ell^2}\frac{1}{1+y}\, .
\ea
We assume that $a$ is positive. Then, one can divide both sides of this equation by $a$. Denote
\be
\hat{b}=b/a\hhh \hat{c}=c/a\hhh
\hat{\ell}=\sqrt{a}\ell\hhh \hat{\mu}=\tilde{\mu}/a\, ,
\ee
Then one has
\be \n{hat}
p+\hat{b}q+\hat{c}v=\frac{1}{\hat{\ell}^2}\frac{1}{1+y}\, .
\ee
The variable $y$ is not changed since
\be
y=\frac{r^{D-1}}{\ell^2  \tilde{\mu}}=\frac{r^{D-1}}{\hat{\ell}^2  \hat{\mu}}\, .
\ee
In what follows, we simply put $a=1$, and we use the equation of the form \eqref{hat}. For simplicity, we also omit ``hats" and write this equation as follows
\be \n{nohat}
p+{b}q+{c}v=\frac{1}{{\ell}^2}\frac{1}{1+y}\, .
\ee

Using \eqref{invars}, one can write this relation as a second order ODE for the master invariant $p$.
This equation, written in explicit form, is
\be\label{eqabc}
\begin{split}
&\CAL{D}p=\dfrac{1}{\ell^2}\dfrac{1}{1+y}\, ,\\
&\CAL{D}=A y^2\dfrac{d^2 }{dy^2}+
B y\dfrac{d }{dy}+C\, ,\\
&A=\dfrac{1}{2}(D-1)^2 c\, ,\\
&B=\dfrac{1}{2}(D-1)[b+(D+2)c]\, ,\\
&C=1+b+c\, .
\end{split}
\ee
We call this equation for the basic curvature invariant $p$ a master equation.
We call a solution of this equation regular if it is smooth for $y\in[0,\infty)$, is finite at $y=0$, and decreases as $1/y$ as $y$ approaches infinity.

\subsection{Solutions}


\subsubsection{Solutions of the homogeneous master equation}

Let us first consider the homogeneous equation
\be \n{HOM}
\CAL{D}p=0\, .
\ee
Let us denote
\be\n{DLL_1}
\begin{split}
&\Delta=\sqrt{(b-c)^2-8c}\, ,\\
&\lambda_{\pm}=\frac{b+3c\pm \Delta}{2(D-1)c}\, .
\end{split}
\ee
In what follows, we shall use the following relations
\be \n{lambda}
\begin{split}
&\lambda_+-\lambda_-=\dfrac{\Delta}{(D-1)c}\, ,\\
&\lambda_+ +\lambda_-=\dfrac{b+3c}{(D-1)c}\, ,\\
&\lambda_+\lambda_-=\dfrac{2(1+b+c)}{(D-1)^2 c}\, .
\end{split}
\ee

For $\Delta>0$,
a general solution of the homogeneous equation \eqref{HOM} is\footnote{
For the degenerate case where $\Delta=0$, $\lambda_+=\lambda_-=\lambda$, and a general solution of the homogeneous equation is
\be\nonumber
p=C_1 y^{-\lambda}+C_2 y^{-\lambda}\ln{y}\, .
\ee
This solution is also non-regular.
}
\be
p=\dfrac{C_+}{y^{\lambda_+}}+\dfrac{C_-}{y^{\lambda_-}}\, .
\ee
Here $C_{\pm}$ are two integration constants.
If $C_{\pm}\ne 0$, then for a solution decreasing at infinity the corresponding power $\lambda_{\pm}$ should be positive and greater than or equal to $1$. However, in such a case this solution is divergent at $y=0$. Thus, the homogeneous equation \eqref{HOM} does not have non-trivial regular solutions.

\subsubsection{A solution of the inhomogeneous master equation}

A solution of the inhomogeneous master equation \eqref{eqabc} can be written in the form\footnote{In what follows, we exclude the degenerate case and always assume that $\Delta>0$.}
\be \n{SOL}
\begin{split}
p&=\CAL{N}\Big[
\Phi_0(y,\lambda_-)
-\Phi_0(y,\lambda_+)\Big]\, ,\\
\CAL{N}&=\frac{2}{(D-1)\ell^2\Delta}\, .
\end{split}
\ee
Here $\Phi_0(y,\lambda)$ is a truncated form of the Lerch transcendent defined in appendix~\ref{AppLerch}, and $\Delta$ is defined in \eqref{DLL_1}. A few typical plots for $p$ in the four-dimensional and five-dimensional cases are presented Figs.~\ref{PLTpD4} and \ref{PLTpD5}.

\begin{figure}[!hbt]
    \centering
\includegraphics[width=0.4\textwidth]{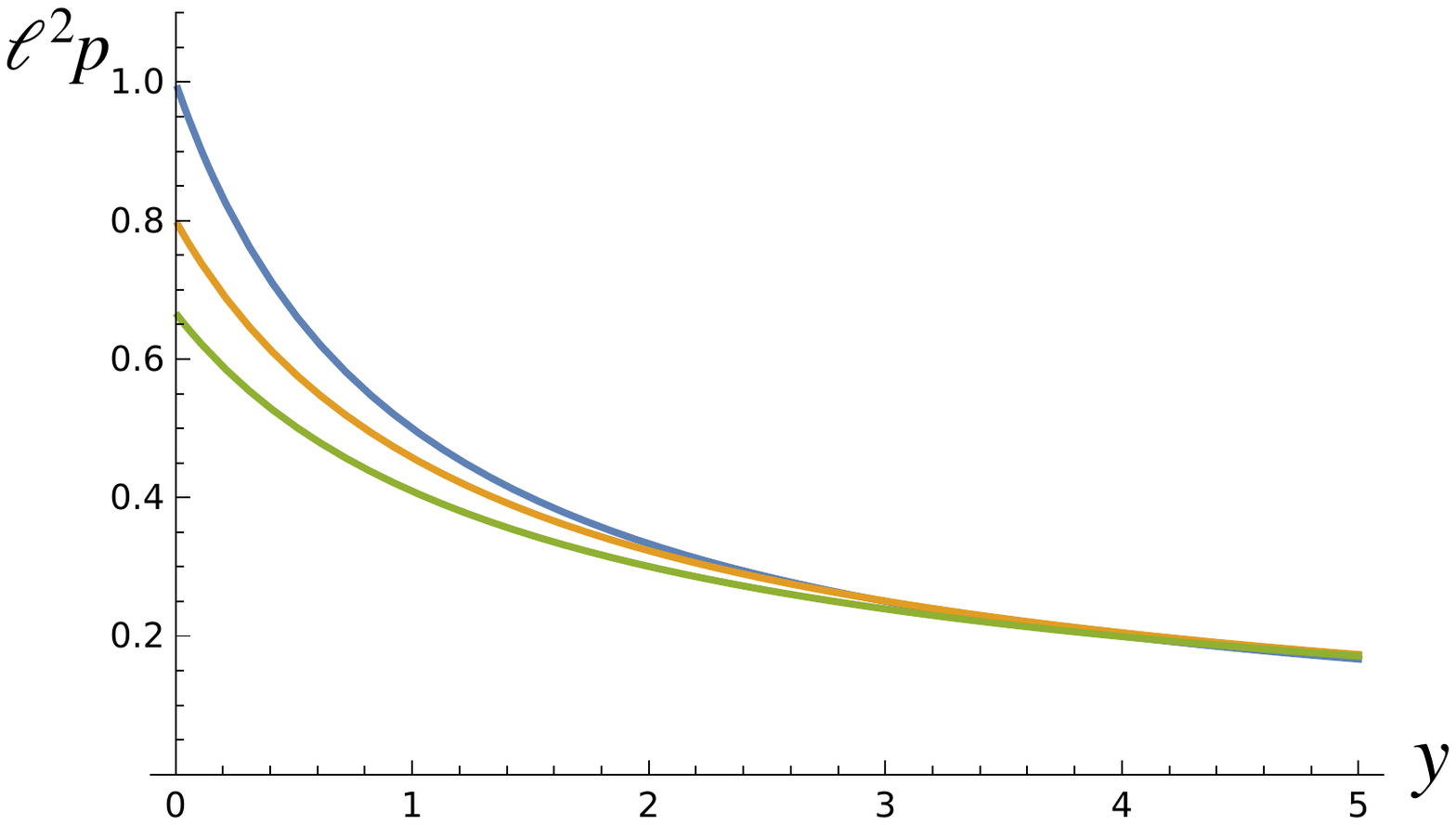}
    \caption{Basic invariant $p$ in $D=4$ dimensions with $c=0$. The cases $b = 0$ (blue), $b=1/4$ (orange) and $b=1/2$ (green) are shown.} \n{PLTpD4}
\end{figure}

\begin{figure}[!hbt]
    \centering
\includegraphics[width=0.4\textwidth]{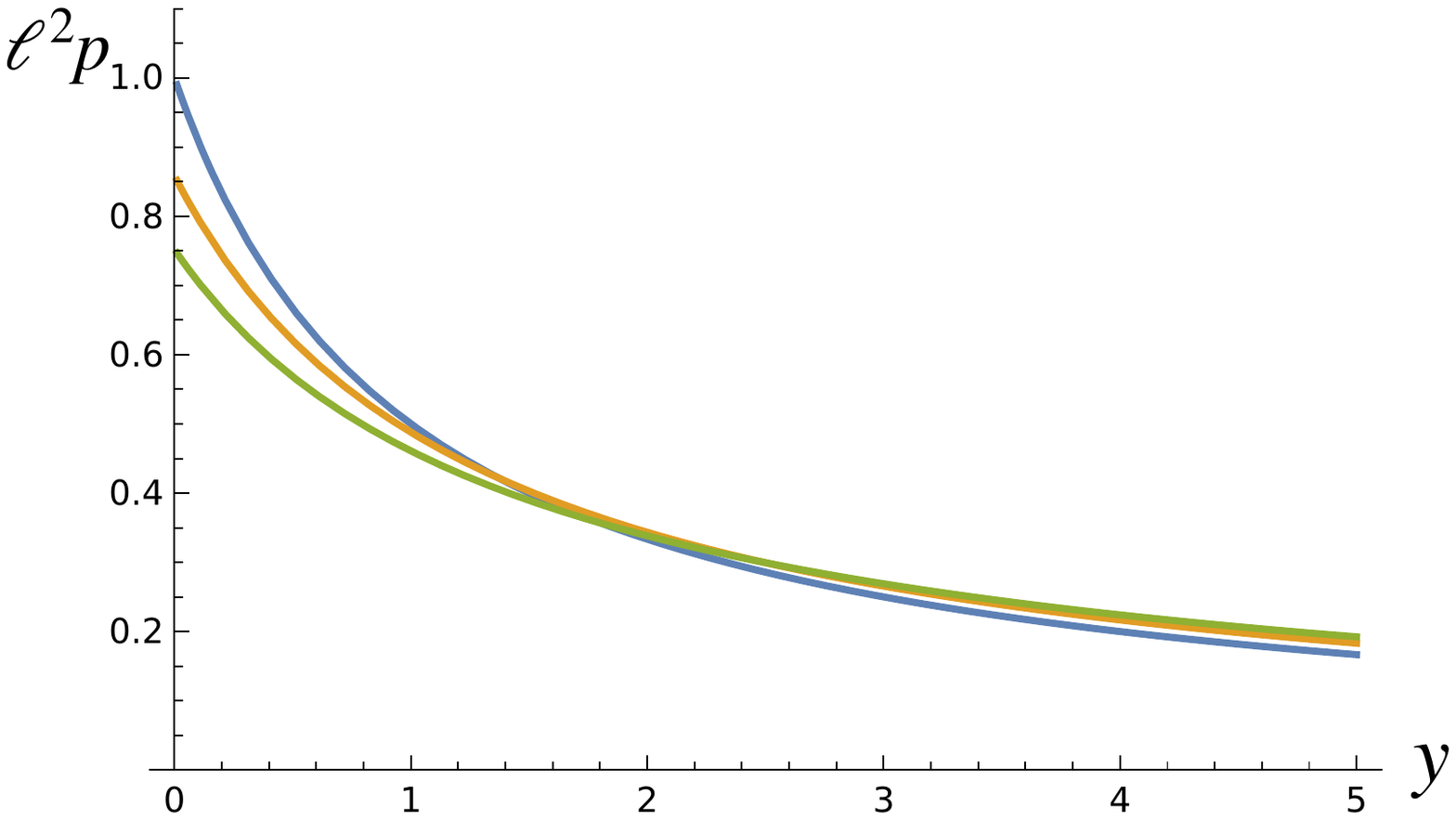}
    \caption{Basic invariant $p$ in $D=5$ dimensions with $c=0$. The cases $b = 0$ (blue), $b=1/6$ (orange) and $b=1/3$ (green) are shown.} \n{PLTpD5}
\end{figure}

At this point we make the following additional assumptions. Namely, we assume that $\lambda_{\pm}$ are real and positive. The condition that they are real allows one to exclude solutions which have oscillatory behavior at infinity. In order to have real $\lambda_{\pm}$, one should impose the condition
\be
(b-c)^2-8c>0 \, .
\ee

Positivity of $\lambda_{\pm}$ makes it possible to use the integral representation for functions $\Phi_0(y,\lambda_{\pm})$, and therefore the corresponding functions have well defined asymptotic behavior at infinity.

Let us first consider the asymptotics of the solution \eqref{SOL} at $y=0$.
Using the relation \eqref{PP0}, one gets
\be
p=\frac{1}{(1+b+c)\ell^2}+O(y)\, .
\ee
This means that the metric function $f$ near $r=0$ has the form
\be
f=1-\frac{r^2}{(1+b+c)\ell^2}+\ldots \, .
\ee
Thus, for a regular black hole, the metric near the origin is de Sitter-like for $1+b+c>0$, and anti-de Sitter-like for $1+b+c<0$. In what follows we assume that the regular black hole has a ``de Sitter core" and focus on the case
\be
1+b+c>0\, .
\ee
Since $\lambda_{\pm}>0$, the third relation in \eqref{lambda} implies that $c>0$. Using the first relation
in \eqref{lambda}, one concludes that
\be
\lambda_+ > \lambda_- \, .
\ee

Using the integral representation \eqref{INT}, one can write
\be
\dfrac{p}{\CAL{N}}=\int_0^{\infty}\dfrac{e^{-\lambda_- x}\big[1-e^{-(\lambda_+-\lambda_-)x} \big] dx}{1+ye^{-x}}\, .
\ee
For $\lambda_+> \lambda_-$, the expression inside the square brackets is positive. This implies that the master scalar invariant $p$ is positive as well.
To obtain a bound on $p$ let us note that $1+ye^{-x}\ge 1$. Thus, putting the denominator equal to 1 one increases the value of the integral. Calculating the obtained integral, one finds the following upper bound for $p$
\be
0<p<\dfrac{1}{\ell^2(1+b+c)}\, .
\ee

Let us now consider the asymptotics of $p$ at infinity. Using \eqref{PPINF}, one has
\be \n{pinf}
p=\CAL{N}\pi \Big[
\dfrac{1}{\sin(\pi\lambda_-)y^{\lambda_-}}-\dfrac{1}{\sin(\pi\lambda_+)y^{\lambda_+}}
\Big]
-\sum_{n=1}^{\infty}\dfrac{B_n}{y^n}\, ,
\ee
where
\be \n{Bn}
B_n=\CAL{N} (-1)^n\Big[\dfrac{1}{\lambda_- -n}-\dfrac{1}{\lambda_+ -n}\Big]\, .
\ee
Calculations give
\be
B_1=\dfrac{1}{Z\ell^2}\, ,\
Z=1-\dfrac{1}{2}(D-3)b +\dfrac{1}{2}(D-3)(D-2)c\, .
\ee

In order to properly reproduce the Schwarzschild-Tangerlini metric at infinity, the term in the square brackets in \eqref{pinf} should decrease faster than $1/y$. To guarantee this, we impose the following condition
\be \n{lak}
\lambda_{-}>k\, ,
\ee
where $k$ is some number larger than 1.
Since $\lambda_+>\lambda_-$, a similar condition for $\lambda_+$ is valid.
Using the definition of $\lambda_-$, one can rewrite \eqref{lak} as follows
\be\n{bcD}
b-\gamma c>\Delta\hh \gamma=2(D-1)k-3\, .
\ee
Hence
\be \n{bc}
b>\gamma c\, .
\ee
For $D\ge 3$ the parameter $\gamma$ is positive. Thus, $b$ should be positive as well. Taking the square of both sides of the inequality \eqref{bcD} and simplifying, one gets
\be \n{lb1}
8-2(\gamma -1)b+(\gamma^2-1) c>0\, .
\ee

Suppose relations \eqref{bc} and \eqref{lb1} are satisfied for $k=1$. Then the term in the square bracket in \eqref{pinf} decreases faster than\footnote{
For the positive integer $l<k$, the first $l$ terms of the series in \eqref{pinf} can be used for an approximation of $p$ at far distance $y$, while the ``correction" due to the ``square bracket contribution" will be small in this domain.
} $1/y$. Then the leading asymptotic of $p$ at infinity is $B_1/y$, and the metric function $f$ has the following form
\be
f=1-\dfrac{\tilde{\mu}}{Zr^{D-3}}+\ldots \, .
\ee
In order to get the proper far-distance asymptotics of the Schwarzschild-Tangherlini metric,
one should put
\be
\tilde{\mu}=Z\mu\, ,
\ee
where $\mu$ is the mass parameter defined in \eqref{muM}.

\begin{figure}[!hbt]
    \centering
\includegraphics[width=0.3\textwidth]{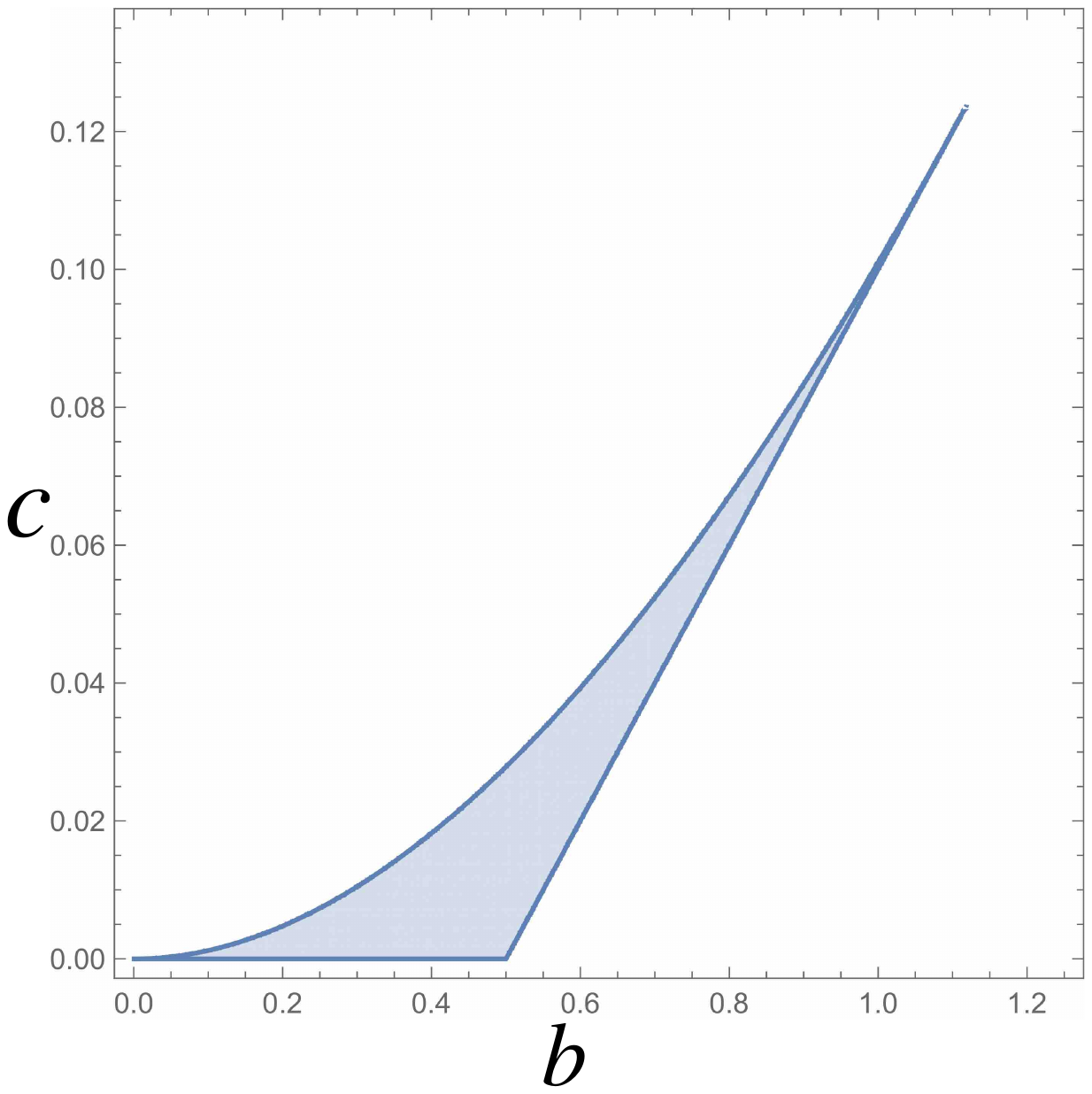}
    \caption{\n{DOMAIN_4} A two-dimensional plane of $(b,c)$ parameters.
    A shaded region in this plane shows allowed values of these parameters for $D=4$ and for the choice $k=2$.  }
\end{figure}

\begin{figure}[!hbt]
    \centering
\includegraphics[width=0.3\textwidth]{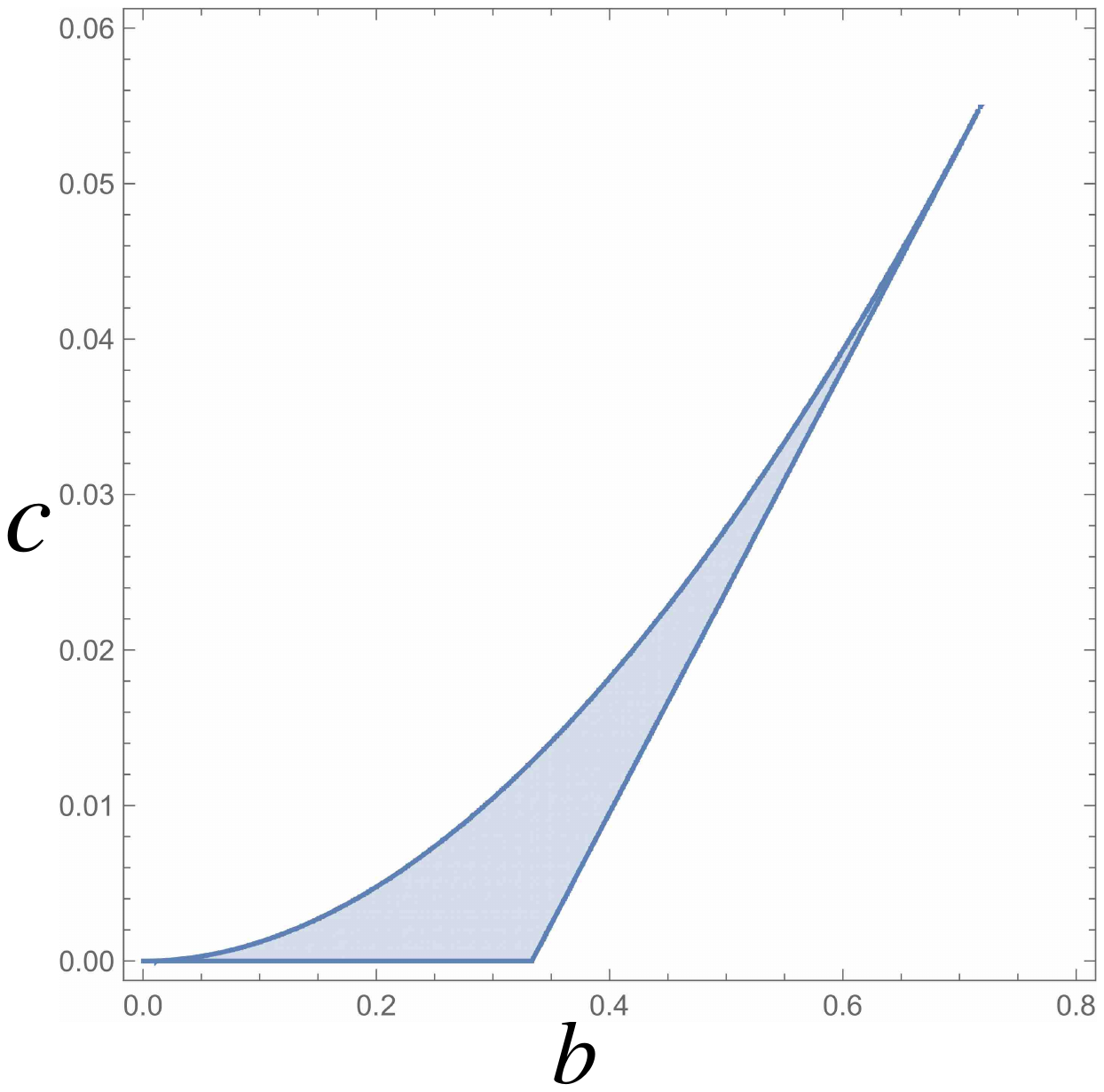}
    \caption{\n{DOMAIN_5}  A two-dimensional plane of $(b,c)$ parameters.
    A shaded region in this plane shows allowed values of these parameters for $D=5$ and for the choice $k=2$.  }
\end{figure}

As we described above, in order to have a regular solution, one needs to impose some restrictions on the model parameters $b$ and $c$.
The domain of the parameter space of $b$ and $c$ depends on the number of dimensions $D$. Shaded regions in Figs.\ref{DOMAIN_4} and \ref{DOMAIN_5} show the allowed values for $b$ and $c$ in dimensions $D=4$ and $D=5$, respectively.  These allowed parameter domains are completely described by the two equations $\lambda_{-} > k$ and $c > 0$ for some $k$ greater than 1.  We can see that our parameter space is bounded by three lines.  The bottom line is $c=0$, the leftmost line is $b = \sqrt{8c} + c$, and the rightmost line is $b = ((D-1)k-1)c + 2/((D-1)k-2)$.  The rightmost line arises from requiring $\lambda_{-} > k$, while the leftmost one arises by necessitating that the argument within the square root of $\Delta$ is positive.

\subsection{Degenerate cases}

\subsubsection{$c=0$ case}
Let us discuss the degenerate case where $c=0$ and $\psi=p+bq$. For this case, one can take limit $c\to 0$ in the above obtained relations. In this limit one has
\be
\Delta=b-\dfrac{(4+b)c}{b}\hh
\lambda_- =\dfrac{2(1+b)}{(D-1)b}\hh
\lambda_+\to \infty\, .
\ee
The integral representation \eqref{INT} for $\Phi_0(y,\lambda)$ shows that in the limit $\lambda\to\infty$, this function vanishes. Thus one has
\be \n{SOLc}
p=\CAL{N}\Phi_0(y,\lambda_-)\hh
\CAL{N}=\frac{2}{(D-1)\ell^2 b}\, .
\ee
Near $r=0$ the metric function $f$ has the form
\be
f=1-\dfrac{r^2}{(1+b)\ell^2}+\ldots \, .
\ee
The condition $1+b>0$ implies that the regular black hole has a ``de Sitter-like core". The condition
\be
4-(\gamma -1)b>0\hh \gamma=2(D-1)k-3
\ee
guarantees that the term $\sim 1/y^{\lambda_-}$ in the solution for $p$ falls faster than the first $k$ terms in its series term. The relation between the integration constant $\tilde{\mu}$ and the mass parameter $\mu$ is of the form
\be
\tilde{\mu}=\big[ 1-\dfrac{1}{2}(D-3)b
\big]\mu\, .
\ee

\subsubsection{$b=c=0$ case}

In this case
\be \n{PAP}
\psi=p\, ,
\ee
and the ODE for the master function reduces to a simple algebraic equation
\be
p=\dfrac{1}{\ell^2}\dfrac{1}{1+y}\, .
\ee
Since in this case $\tilde{\mu}=\mu$, one can write this relation in the following form
\be \n{pmu}
p=\dfrac{{\mu}}{r^{D-1}+\ell^2{\mu} }\, .
\ee
Written in this form, the expression for $p$ reproduces the master invariant for the Hayward metric. At the same time, the model reduces to the special case of quasi-topological gravity.

Let us note that the Schwarzschild-Tangherlini metric is recovered by simply putting $\ell=0$ in \eqref{pmu}.

\section{Regular black hole solutions and their properties}\label{sec5}

\subsection{Basic curvature invariants}

The solution of the inhomogeneous equation for the invariant $p$ is given by \eqref{SOL}, while the invariants $q$ and $v$ are given by
\ba
q=&\CAL{N}\Big[
\Phi_0(y,\lambda_-)-\Phi_0(y,\lambda_+)\\
&-\frac{D-1}{2}\big(\Phi_1(y,\lambda_-)-\Phi_1(y,\lambda_+)\big)\Big] ,\\
v=&\CAL{N}\Big[
\Phi_0(y,\lambda_-)-\Phi_0(y,\lambda_+)\\
&-\frac{(D-1)(D+2)}{2}\big(\Phi_1(y,\lambda_-)-\Phi_1(y,\lambda_+)\big)
\\
&+\frac{(D-1)^2}{2}\big(\Phi_2(y,\lambda_-)-\Phi_2(y,\lambda_+)\big)\Big] .
\ea

Since functions $\Phi_i(y,\lambda)$ for $\lambda>0$ are bounded (see Appendix~C), this property is also valid for other basic curvature invariants, as well as polynomials constructed from them.

\begin{figure}[!hbt]
    \centering
\includegraphics[width=0.40\textwidth]{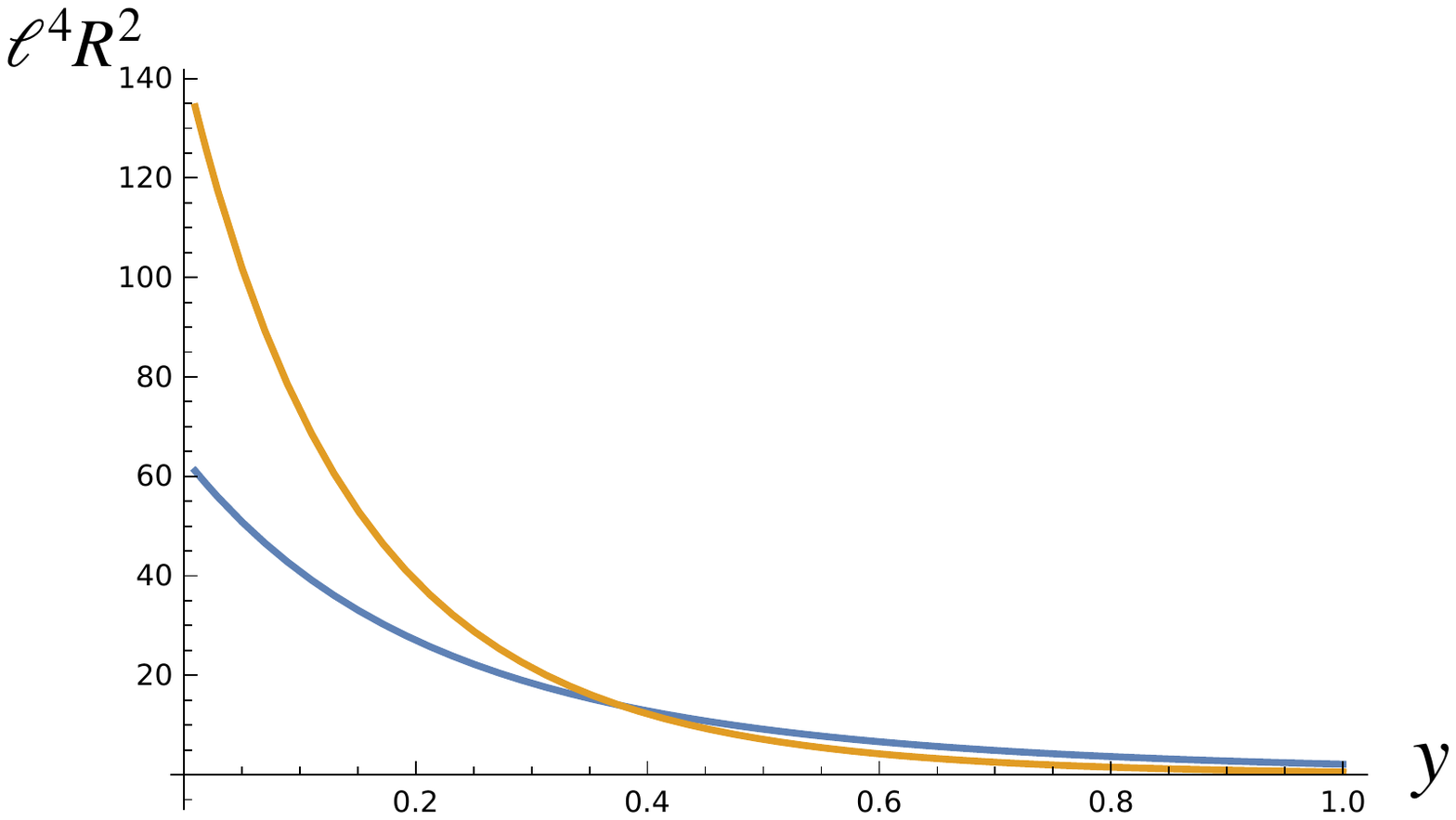}
\caption{Dimensionless curvature invariant $\ell^{4}R^{2}$ as a function of $y$ in dimension $D=4$ with $c=0$ and $b=0$ (orange), $b=b_m$ (blue).}\n{PLTR2}
\end{figure}
\begin{figure}[!hbt]
    \centering
\includegraphics[width=0.40\textwidth]{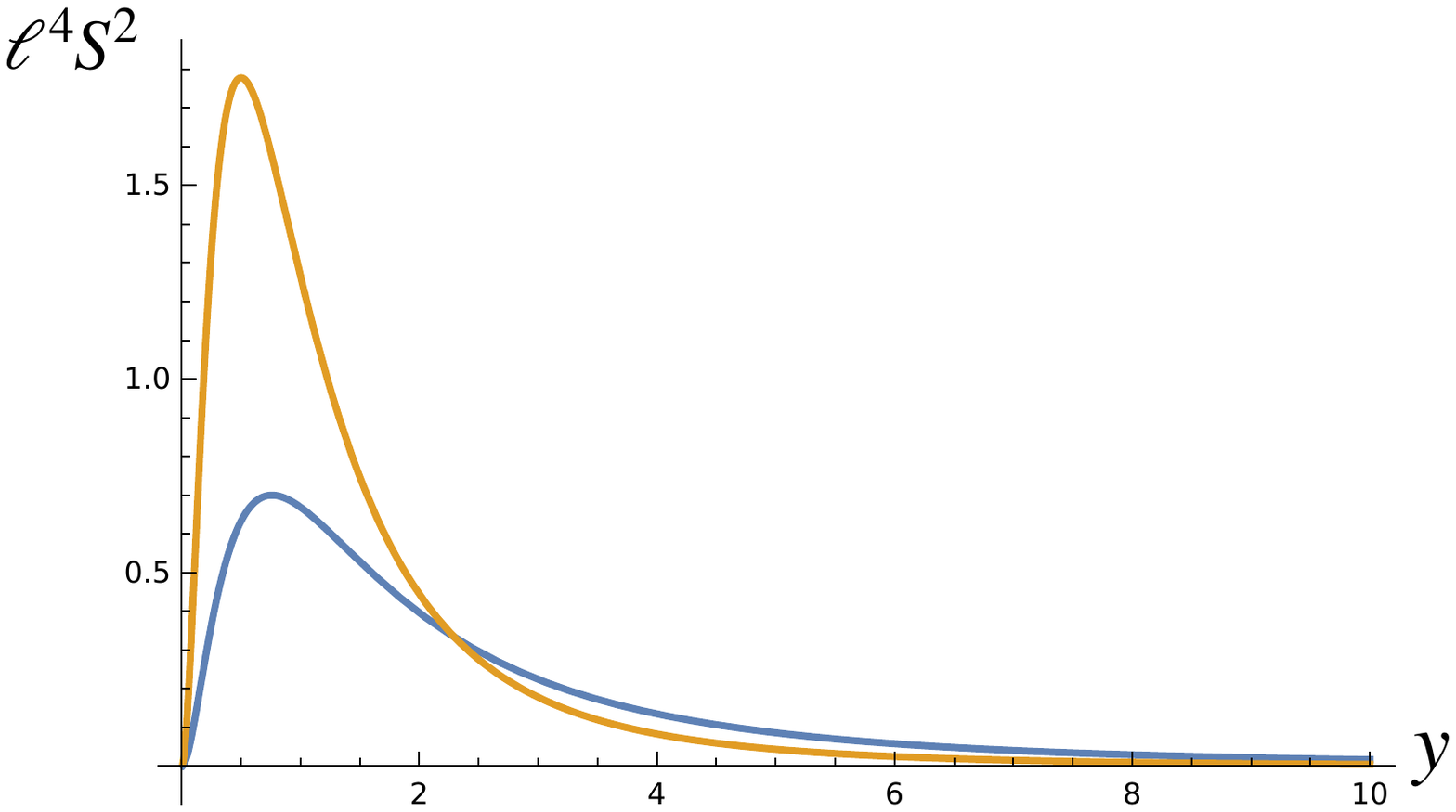}
 \caption{Dimensionless curvature invariant $\ell^{4}S_{\mu\nu}^{2}$ as a function of $y$ in dimension $D=4$ with $c=0$ and $b=0$ (orange), $b=b_m$ (blue).}\n{PLTS2}
\end{figure}
\begin{figure}[!hbt]
    \centering
\includegraphics[width=0.40\textwidth]{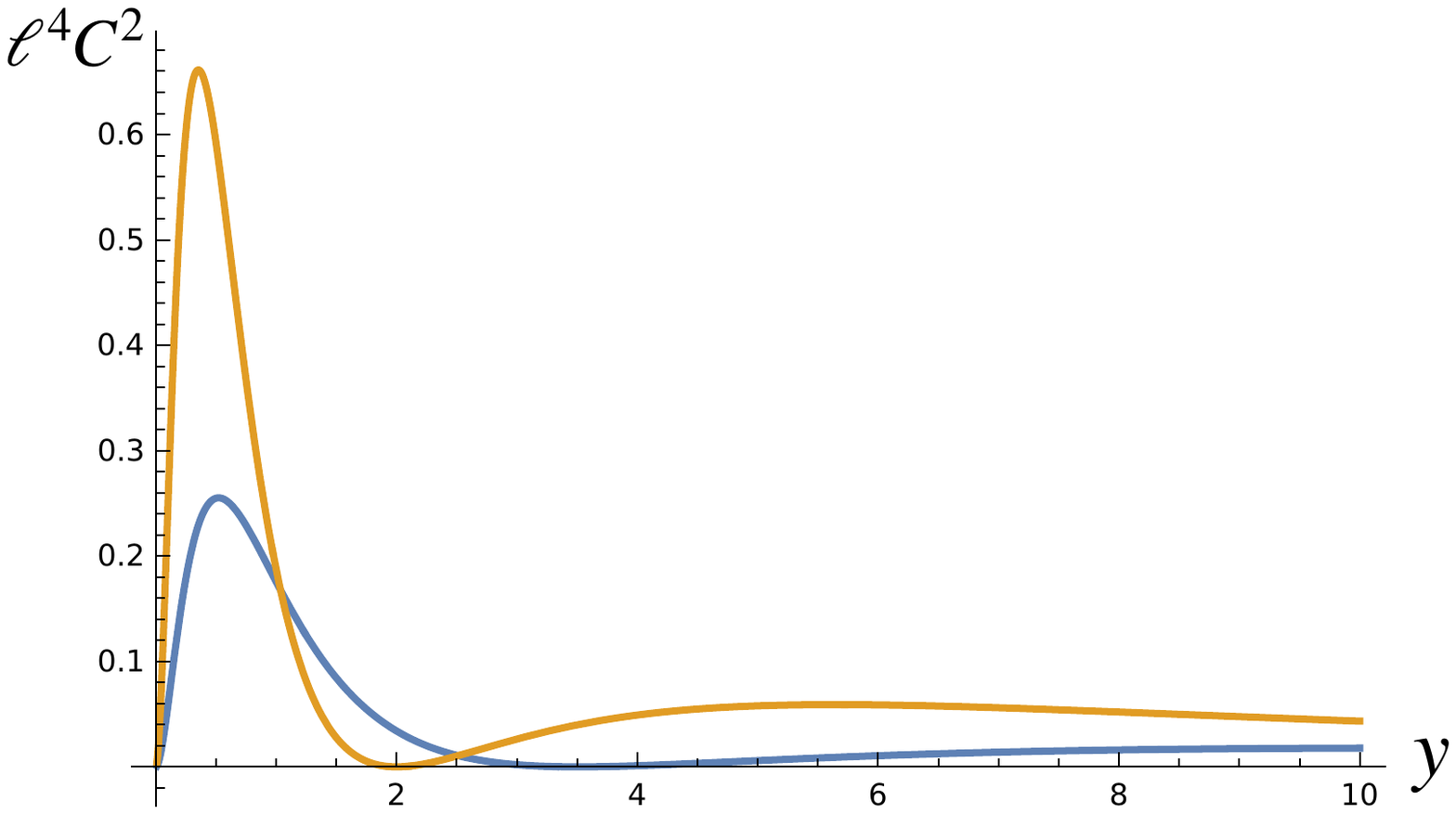}
\caption{Dimensionless curvature invariant $\ell^{4}C^{2}$ as a function of $y$ in dimension $D=4$ with $c=0$ and $b=0$ (orange), $b=b_m$ (blue).}\n{PLTC2}
\end{figure}

Let us note that the solution \eqref{SOL} , in addition to depending on the parameters of the model $b$ and $c$, depends on the number of dimensions $D$. In order to illustrate basic properties of this solution we need to make a  special choice of these parameters. We first choose to look at the simplest case in terms of dimensions and set $D=4$. We also put $c=0$, and choose the values of the parameter $b$ in the allowed domain, between $b=0$ (the Hayward solution) and $b=b_m$, where $b_m$ is the maximal value of $b$ allowed by constraints:
$b_m=1/2$ for $D=4$ and $b_m=1/3$ for $D=5$.

To illustrate properties of the curvature invariants, we include three plots Figs.~\ref{PLTR2},\ref{PLTS2},\ref{PLTC2} that follow the special choice of our parameters outlined above and demonstrate
how the curvature invariants $R^2$, $S_{\alpha\beta}S^{\alpha\beta}$, and $C_{\alpha\beta\gamma\delta}C^{\alpha\beta\gamma\delta}$ depend on $y$.
We use the notation $S_{\mu\nu}=R_{\mu\nu}-g_{\mu\nu}R/D$ for traceless part of the Ricci tensor

\subsection{Metric}

The metric function $f$ can be presented in the form
\be\n{mety}
\begin{split}
&f=1-F_D \Psi(y)\hh F_D=\Big[
\dfrac{\tilde{\mu}}{l^{D-3}}
\Big]^{2/(D-1)}\, ,\\
&\Psi(y)=\ell^2 y^{2/(D-1)}p(y)\, .
\end{split}
\ee
Here $y$, $F_D$, and $\Psi(y)$ are all dimensionless quantities.
Let us emphasize that this form of the metric function $f$ is a consequence of the scaling property of our model discussed in subsection~\ref{SS_3B}. This form is very useful since the function $\Psi(y)$ depends only of the number of dimensions and parameters $b$ and $c$ of the model, while $F_D$ keeps track of the mass parameter, which specifies a solution.

The function $f$ decreases from 1 at $y=0$, reaches a minimal value at the point where $d\Psi/dy=0$, and after this monotonically increases towards 1.
For a sufficiently large value of the dimensionless parameter $F_D$, the metric has two horizons and the solution describes a regular black hole.
The equation $f=0$ gives
\be\label{PhiH}
\Psi(y_H)=\frac{1}{F_D}\, .
\ee
For small $F_D$, the metric does not have a horizon. There exists
a critical value $F_D^*$ at which the inner and outer horizons merge.
The corresponding critical values $y_*$ and $F_D^*$ for critical solution
can be found by solving the following two equations
\be\n{CRIT}
\Psi(y)=\frac{1}{F_D}\hh \dfrac{d\Psi}{dy}=0\, .
\ee
The second of these equations determines the critical value of the coordinate $y=y_*$. It can be written in the form
\ba\n{CRIT2}
\dfrac{2}{D-1}&\big[\Phi_0(y,\lambda_-) -
\Phi_0(y,\lambda_+)\big] \\
+y\dfrac{d}{dy}
&\big[\Phi_0(y,\lambda_-) -
\Phi_0(y,\lambda_+)\big]=0\, .
\ea
Using the integral representation for $\Phi_0(y,\lambda)$, one can show that
\be \n{CRIT_1}
y\dfrac{d}{dy}\Phi_0(y,\lambda)=\dfrac{1}{1+y}-\lambda \Phi_0(y,\lambda)\, .
\ee
Using this relation and \eqref{CRIT2}, one obtains the following equation for the critical parameter $y_*$
\be \n{CRIT_2}
\big(\lambda_+ -\dfrac{2}{D-1}\big)
\Phi_0(y_*,\lambda_+)=\big(\lambda_- -\dfrac{2}{D-1}\big)
\Phi_0(y_*,\lambda_-).
\ee

For the special case where the parameter $c=0$, similar calculations give the following equation for $y_*$
\be
\dfrac{2}{(D-1)b}\Phi_0(y_*,\lambda_-)=\dfrac{1}{1+y_*}\, .
\ee

The critical value of the black hole's mass can be found by means of the following relation
\be
F_D^*=\dfrac{1}{\Psi(y_*)}\, .
\ee
In the case where $c=0$, this equation takes a simpler form
\be
F_D^* = y_*^{-\frac{2}{D-1}}(1+y_*) .
\ee

\begin{table}[h!]
\hfill
\begin{tabular}{ |l|l| }
 \hline
 \multicolumn{2}{| c |}{$(y_*,F_D^*)$ in $D=4$}\\
 \hline
 (2.000, 1.890) & $b = 0$\\
 (2.922, 1.919) & $b = b_{m}/2$ \\
 (4.116, 1.992) & $b = b_{m}$  \\
 \hline
\end{tabular}
\hfill
\begin{tabular}{ |l|l| }
 \hline
 \multicolumn{2}{| c |}{$(y_*,F_D^*)$ in $D=5$}\\
 \hline
(1.000, 2.000)& $b = 0$ \\
(1.378, 2.026)  & $b = b_{m}/2$ \\
(1.817, 2.089)  & $b = b_{m}$   \\
 \hline
\end{tabular}
\hspace{1cm}
\caption{The left table gives critical values of $y_*$ and $F_D^*$ in $D=4$ spacetime dimensions for $c=0$ and different values of the parameter $b$, where $b_m=1/2$. The right table gives similar quantities for $D=5$, where $b_m=1/3$.}
\end{table}

\begin{figure}[!hbt]
    \centering
\includegraphics[width=0.45\textwidth]{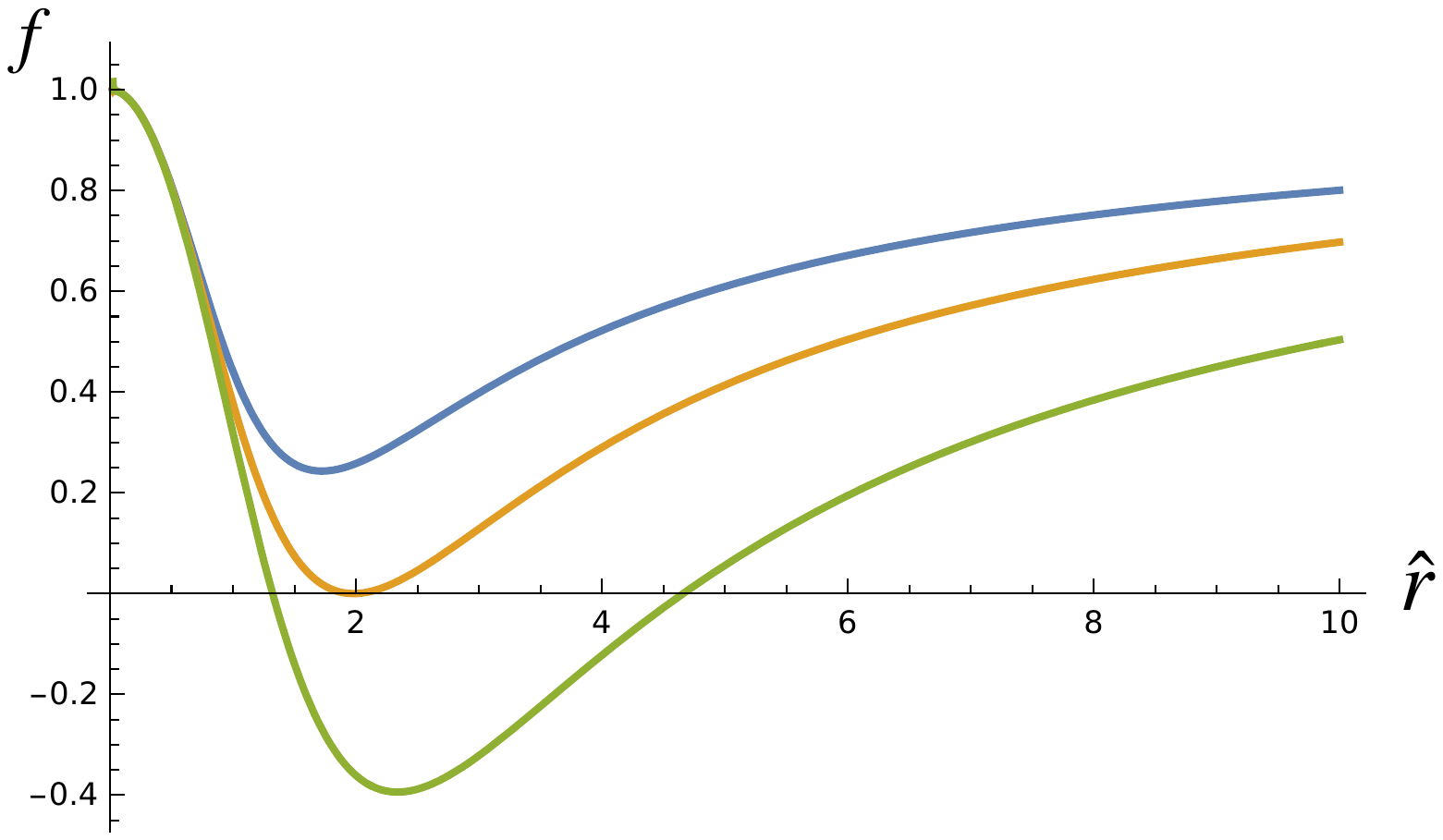}
    \caption{Metric function $f$ in terms of dimensionless parameter $\hat{r}$ in $D=4$ dimensions for $b=b_m/2$, $c=0$ with $\hat{r}_\ins{g}=2$ (blue), $ \hat{r}_\ins{g}\approx 3.03$ (orange), and $ \hat{r}_\ins{g}=5 $ (green). }\label{PLTfD4}
\end{figure}

\begin{figure}[!hbt]
    \centering
\includegraphics[width=0.45\textwidth]{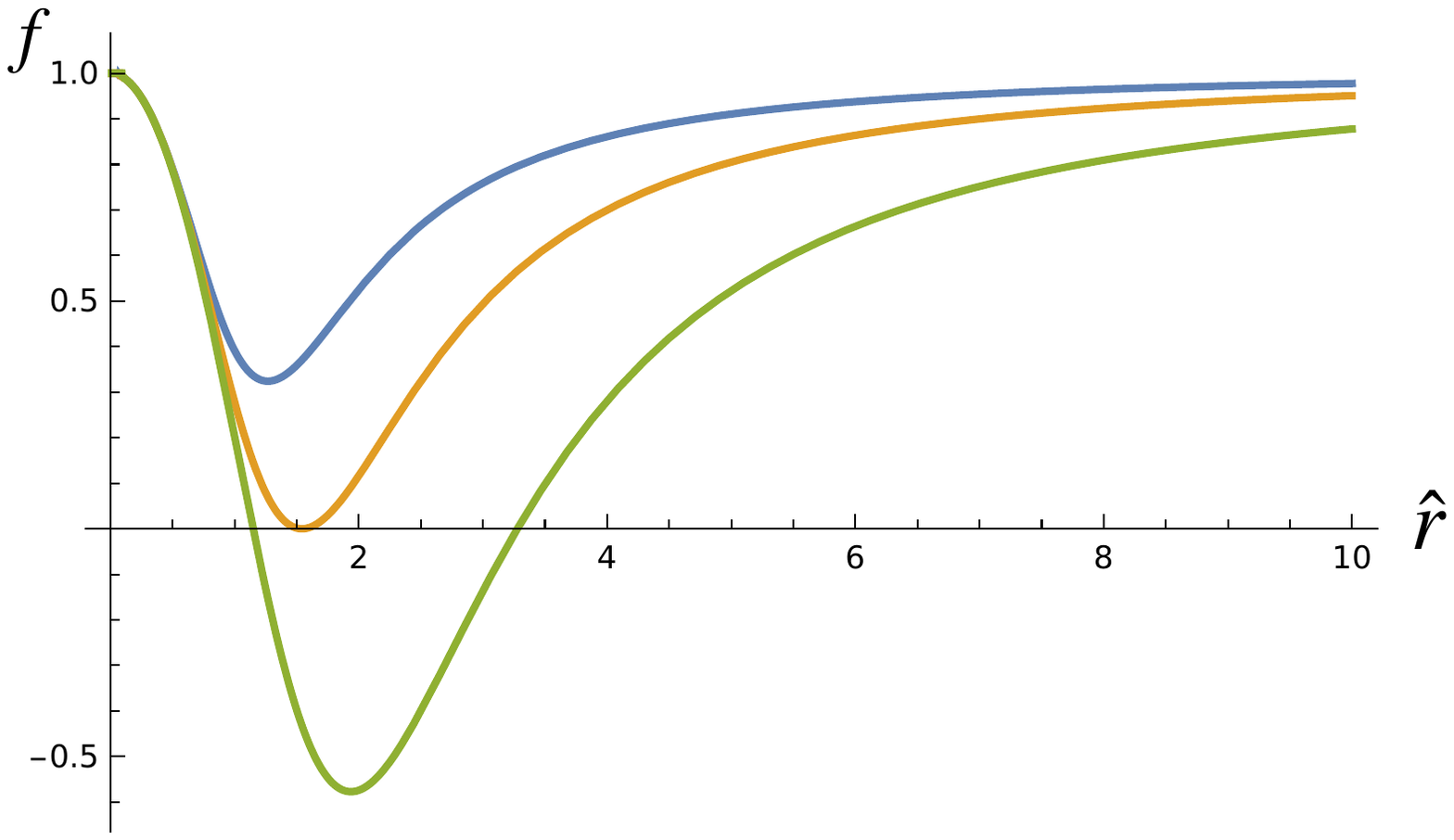}
  \caption{Metric function $f$ in terms of dimensionless parameter $\hat{r}$ in $D=5$ dimensions for $b=b_m/2$, $c=0$ with $\hat{r}_\ins{g}=1.5$ (blue), $ \hat{r}_\ins{g}\approx 2.22$ (orange), and $ \hat{r}_\ins{g}=3.5 $ (green). }\label{PLTfD5}
\end{figure}

Let us remember that the metric function $f$ enters the metric in $(t,r)$ coordinates. For this reason, it is also instructive to present the expression \eqref{mety} of $f$ as a function of the $r$-coordinate as well. For this purpose, we use the following relations
\be
\begin{split}
&\tilde{\mu}=Z\mu=Zr_\ins{g}^{D-3}\, ,\\
&Z=1-\dfrac{1}{2}(D-3)b +\dfrac{1}{2}(D-3)(D-2)c\, ,\\
&\hat{r}=r/\ell\hh \hat{r}_\ins{g}=r_\ins{g}/\ell\, ,\\
& F_D=Z^{{2}/{D-1}}
 \hat{r}_\ins{g}^{{2(D-3)}/({D-1})}\, ,\\
&y=\dfrac{1}{Z}\dfrac{\hat{r}^{D-1}}{\hat{r}_\ins{g}^{D-3}}\, .
\end{split}
\ee
Figs.~\ref{PLTfD4} and \ref{PLTfD5} illustrate the behavior of the function $f(\hat{r})$ in four- and five-dimensional spacetimes, respectively.

\subsection{Thermodynamic parameters}

\subsubsection{General relations}

The Hawking temperature $T_\ins{H}$ of a black hole is related to the surface gravity $\kappa$ on the event horizon as follows
\ba
T_\ins{H}=\hbar\frac{\kappa}{2\pi} .
\ea
We work in the system of units where the speed of light $c=1$ and the Boltzmann constant $k_\ins{B}=1$, but we keep $\hbar$ intact.
The surface gravity for the geometry \eqref{ds} expressed in $(t,r)$-coordinates and measured on the static solution with $N=1$ is given by the formula
\ba
\kappa=\frac{1}{2}\Big|\frac{df}{dr}\Big|_\ins{H}=r_\ins{H} |q|_\ins{H} .
\ea
Note that $r_\ins{H}$ is the radius of the horizon of our model, defined by the equation $f=0$. It is different from the parameter $r_\ins{g}$ which is related to the asymptotic mass of the black hole. In Einstein's theory, for a pure Schwarzschild-Tangherlini solution they coincide.

The variation of the energy $dE=dM$ and the variation of the entropy $dS$ are related by the second law of thermodynamics
\be
dM=TdS .
\ee
Taking into account \eqref{muM}, we get
\be\label{dS}
dS=\frac{2\pi B}{\hbar}\frac{1}{\kappa}d\mu .
\ee
On the event horizon, the condition $f(r_\ins{H})=0$ makes it possible to find $\mu$ as a function of $r_\ins{H}$.
Substituting functions of gravitational radius $\kappa(r_\ins{H})$ and $\mu(r_\ins{H})$ into \eqref{dS} and integrating this equation, we get
\ba\n{ENT}
S=S_*+\frac{2\pi B}{\hbar}\int_{r_*}^{{r_\ins{H}}}d\xi \frac{1}{\kappa(\xi)}\frac{d\mu(\xi)}{d\xi} .
\ea
Here $r_*$ is the gravitational radius of the minimal-mass black hole. The latter is defined by the condition that horizon is degenerate, i.e., $r_+=r_-$, which corresponds to the requirement
that $\kappa(r_*)=0$.

\subsubsection{The Hayward solution}

For the general solution \eqref{SOL}, the integral in \eqref{ENT} cannot be evaluated in an explicit form, and if needed, one can calculate it numerically. However, for a simple model with  $a=1,~b=0,~c=0$ one can use \eqref{ENT} to get the analytic expression for the entropy $S$. In this case, the solution coincides with the Hayward metric,
and the metric function $f$ is
\be
f=1-\dfrac{{\mu}r^2}{r^{D-1}+\ell^2{\mu} }.
\ee

Thus we have
\be
\mu(r_\ins{H})=\frac{r_\ins{H}^{D-1}}{r_\ins{H}^2-\ell^2}\,,
\ee
and
\be
\kappa(r_\ins{H})=\frac{(D-3)r_\ins{H}^2-(D-1)\ell^2}{2r_\ins{H}^3}\,.
\ee
The horizon radius of the minimal black hole is
\be
r_*=\sqrt{\frac{D-1}{D-3}}\ell\,,
\ee
and its minimal parameter $\mu_*$ is given by
\be
\mu_*=\frac{(D-1)^{(D-2)/2}}{2(D-3)^{(D-3)/2}}\ell^{D-3}\,.
\ee
For the variation of entropy, we obtain
\be
dS=\frac{4\pi B}{\hbar}\frac{r_\ins{H}^{D+1}}{(r_\ins{H}^2-\ell^2)^2}\,dr_\ins{H}\,.
\ee
The value of $S_*$ is similar to the entropy of an extremally charged black hole. It is not fixed by this approach, but may be determined from other considerations \cite{Bueno:2022res}, like Wald's method.
Let us introduce the Planck length $\ell_\ins{Pl}$ via the equality
$\ell_\ins{Pl}^{D-2}=\hbar\varkappa/(8\pi)$.
Then in four-dimensions we get
\ba\label{Entr}
S=&S_*+\frac{\pi}{2\ell_\ins{Pl}^2(r_\ins{H}^2-\ell^2)}\Big[2r_\ins{H}^4-7\ell^2r_\ins{H}^2+3\ell^4 \\
&+4\ell^2( r_\ins{H}^2-\ell^2 )\Big(\ln\big(\frac{r_\ins{H}^2}{\ell^2}-1\big)-\ln 2\Big)
\Big]\,.
\ea
At small $\ell\ll r_\ins{H}$, we have
\be\label{Entr0}
S=S_*+\frac{\pi r_\ins{H}^2}{\ell_\ins{Pl}^2}+\frac{\pi\ell^2}{\ell_\ins{Pl}^2}
\Big[4\ln\Big(\frac{r_\ins{H}}{\ell}\Big)-2\ln 2 -\frac{5}{2}\Big]+O(\ell^4) \,.
\ee
In the limit $\ell\ll r_\ins{H}$, the relation between $r_\ins{H}$ and $r_\ins{g}$ in $D$ dimensions reads
\be
r_\ins{H}=r_\ins{g}-\frac{\ell^2}{(D-3)r_\ins{g}}+O(\ell^4) \,.
\ee
The leading term $\pi r_\ins{H}^2/\ell_\ins{Pl}^2=\CAL{A}/(4 \ell_\ins{Pl}^2)$ in  Eq.\eqref{Entr0} reproduces the Bekenstein-Hawking entropy term, which is proportional to the area $\CAL{A}$ of the horizon.


\section{Discussion}\label{discussion}

In this paper we derived a wide set of new metrics describing regular spherically symmetric black holes. These metrics are solutions of a specially constructed two-dimensional dilaton gravity model with the action \eqref{SLL}, which is non-linearly dependent on the basic curvature invariants. The Lagrangian density of this action is
$\CAL{L}=-\CAL{W}h$, where $\CAL{W}$ and $h$ are defined by \eqref{WWWW} and \eqref{hpqvu}. For $h=p$, this model reproduces the spherically reduced Einstein-Hilbert action, while for an arbitrary function $h=h(p)$ it coincides with the action of the quasi-topological model \cite{Bueno:2024dgm}. For a special choice
\be
h(p)=\dfrac{p}{1-\ell^2 p}\, ,
\ee
where $\ell$ is some ``fundamental" length,
the solution of reproduces the Hayward metric.
Remarkably, as was proved in section \ref{SS_3C}, the covariant form of the $D$-dimensional action can be reconstructed for any arbitrary function $h(p)$. However, for the more general function $h(p,q,u,v)$, the possibility of upgrading the $2D$ dilaton action to a $D$-dimensional covariant action is not so evident and requires further analysis.

We considered a general case, where $h$ is an arbitrary function of all the basic curvature invariants of the metric \eqref{ds}, and demonstrated
that solutions of such a theory have a remarkable property. There exists such a choice of a new coordinate $y$ given by \eqref{yyy}, for which the invariant $p$ has the form $\ell^{-2} F(y)$. For regular black holes, the function $F$ is regular and bounded $|F(y)|<B$ on the interval $y\in [0,\infty)$, where $B$ is some dimensionless constant,  and $F\sim 1/y$ at infinity. We demonstrated that the other basic curvature invariants $q$, $u$, and $v$ have similar properties, and hence
the polynomial curvature invariants are also bounded. This implies that such a model satisfies Markov's limiting curvature condition.

After this, we focused on a class of non-linear in curvature models with a special choice of the ``form factor" $h$
\be
h(\psi)=\dfrac{\psi}{1-\ell^2 \psi}\hh
\psi=p+bq+cv\, .
\ee
We demonstrated that for such a choice, the corresponding field equation for the metric function $f$ is an inhomogeneous second order linear ordinary differential equation. We were able to solve this equation explicitly and describe a range of the parameters $(b,c)$ for which the corresponding solutions are regular. Using this result, we obtained a new wide class of regular black hole solutions and discussed their properties.

As expected, the obtained regular solutions have a horizon only if their mass parameter $\tilde{\mu}$ exceeds some critical value $\tilde{\mu}_*$. In other words, there exists a mass gap for the formation of such black holes \cite{Frolov:2015bta}. Let $r_*$ be a corresponding critical radius. Then the two dimensionless critical parameters $r_*/\ell$ and $\tilde{\mu}_*/\ell^{D-3}$ are defined as a solution of two equations $f=0,~f'=0$.
An interesting property of the model discussed in this paper is the following. There exists a special dimensionless combination $y_*=r_*^{D-1}/(\ell^2 \tilde{\mu}_*)$ of these two critical parameters which is determined by a single equation (see \eqref{CRIT_1} and \eqref{CRIT_2}). In this sense
the critical solution is universal. This property is a consequence of the scaling property of the model.

Let us emphasize once again that we did not require that the model we discussed in the paper was obtained by the spherical reduction of a covariant $D$-dimensional action. Rather, we used it as a special two-dimensional dilaton gravity model which ``generates" regular black hole solutions. We demonstrated that this model allows solutions with $N=1$, and we focused on studying these solutions. In this paper we did nor answer the question of whether there exist regular black hole solutions with non-constant $N$. There are also questions that remain about the stability of the obtained solutions. Certainly, it would be interesting to address these questions as well as to check whether this type of models are able to ``cure" the mass inflation ``disease".

\appendix


\section{Curvature invariants}\label{AppInv}

All other  ocal scalar invariants constructed from the curvature can be expressed as a function of the four functions $p$,$q$, $u$, and $v$, whose explicit forms for the metric (\ref{ds}) are given in (\ref{pquv0}). For example, the Ricci scalar has the form
\ba\label{Ra}
R=&\big[v+(D-2)q\big]+\big[v+(D-2)u\big]\\&+(D-2)\big[(D-3)p+q+u\big]\,.
\ea
Higher order in curvature invariants are
\ba
R_{\mu}^{\nu} R_{\nu}^{\mu}
=&\big[v+(D-2)q\big]^2+\big[v+(D-2)u\big]^2\\
&+(D-2)\big[(D-3)p+q+u\big]^2\,,
\ea
\ba
R_{\mu}^{\nu} R_{\nu}^{\sigma}R_{\sigma}^{\mu}
=&\big[v+(D-2)q\big]^3+\big[v+(D-2)u\big]^3\\
&+(D-2)\big[(D-3)p+q+u\big]^3\,,
\ea
\ba
C_{\alpha\beta\mu\nu} C^{\alpha\beta\mu\nu}=&\frac{4(D-3)}{D-1}(p-q-u+v)^2\, .
\ea

If needed, one could reverse these relations and express invariants $p$,$q$, $u$, and $v$ in terms of $R$, $C_{\alpha\beta\mu\nu} C^{\alpha\beta\mu\nu}$, $R_{\mu}^{\nu} R_{\nu}^{\mu}$ , and $R_{\mu}^{\nu} R_{\nu}^{\sigma}R_{\sigma}^{\mu}$. However, this representation requires solving a cubic algebraic equation, and is quite involved.


\section{Properties of the Lerch transcendent}\label{AppLerch}

In this appendix we collect main formulas concerning the Lerch transcendent. Basic information about these functions can be found in \cite{bateman1953higher,olver2010nist}.
For asymptotics, and discussion of properties of the Lerch transcendent, see also \cite{lerch,FERREIRA2004210,Lagarias_2016}.

Let $z$ be a complex variable. Then the Lerch transcendent for $|z|<1$ is defined as
\be
\Phi\big(z,s,\lambda\big)=\sum_{k=0}^{\infty}\frac{z^k}{(\lambda+k)^s}\, .
\ee
If $s$ is a positive integer, then $\lambda\ne 0,-1,-2,\ldots$.
For other values of the complex argument $z$, $\Phi\big(z,s,\lambda\big)$ is defined by analytic continuation.
The following integral
\be
\Phi(z,s,\lambda)=\dfrac{1}{\Gamma(s)}\int_0^{\infty}\dfrac{x^{s-1}e^{-\lambda x}}{1-z e^{-x}}\,,
\ee
defines the Lerch transcendent for $\Re{\lambda}>0$ and $\Re{s}>0$ as an analytic function of the complex variable $z$ on the complex plane, with a cut on the real axis at $z\in  [1,\infty)$.

For every positive integer $m$, there is a useful property
\ba
\Phi\big(z,s,\lambda\big)=z^m\Phi\big(z,s,\lambda+m\big)+\sum_{k=0}^{m-1}\frac{z^k}{(\lambda+k)^s}\,.
\ea

In the main text, we use a special version of the Lerch transcendent for the case where $s=1$ and its argument $z$ is real and negative.
We denote this function as $\Phi_0(y,\lambda)$ (see Fig.~\ref{PLTLerch} )
\be
\Phi_0(y,\lambda)=\Phi(-y,1,\lambda)\, .
\ee
Its integral form, which is valid for $\Re{\lambda}>0$, is
\be\n{INT}
\Phi_0(y,\lambda)=\int_0^{\infty}\dfrac{e^{-\lambda x} dx}{1+y e^{-x}}\, .
\ee

\begin{figure}[!hbt]
    \centering
\includegraphics[width=0.4\textwidth]{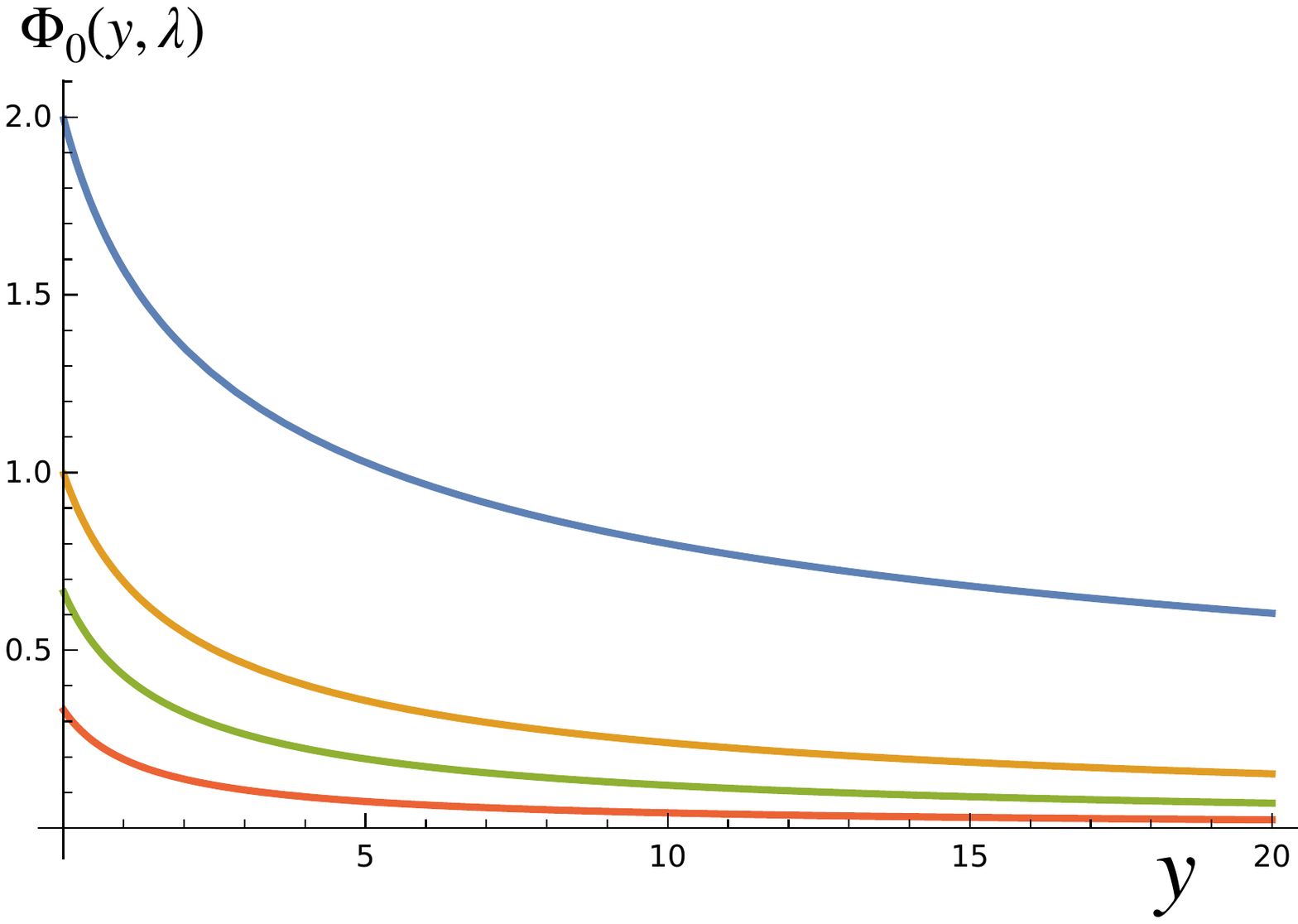}
     \caption{Plots of the function $\Phi_{0}(y,\lambda)$ for the values of parameter $\lambda = 0.5$ (blue), $1$ (orange), $1.5$ (green), $3$ (red).}\n{PLTLerch}
\end{figure}

For $\Re\lambda>0$, $\Phi_0(y,\lambda)$ has the following asymptotics. For $y\to 0$, one has
\be \n{PP0}
\Phi_0(y,\lambda)=\frac{1}{\lambda}-\frac{y}{\lambda+1}+O(y^2)\,.
\ee
For $y\to\infty$, one has (see \cite{lerch})
\ba\n{PPINF}
\Phi_0(y,\lambda)=\frac{\pi}{\sin(\pi\lambda)}\frac{1}{y^\lambda}
-\sum_{n=1}^{\infty}\frac{(-1)^n}{(\lambda-n) y^n}\,.
\ea

In the main text, we use also the following functions obtained from $\Phi_0(y,\lambda)$ via differentiation
\be \n{PPP}
\begin{split}
\Phi_1(y,\lambda)&=-y\dfrac{d}{dy}\Phi_0(y,\lambda)\\
&=\lambda\Phi_0(y,\lambda)-\frac{1}{1+y}     \, ,\\
\Phi_2(y,\lambda)&=y^2\dfrac{d^2}{dy^2}\Phi_0(y,\lambda)\\
&=\lambda(\lambda+1)\Phi_0(y,\lambda)-\frac{1+\lambda}{1+y}-\frac{y}{(1+y)^2}\, .
\end{split}
\ee
Let us denote
\be
\begin{split}
h_0&=\dfrac{1}{1+w}\hh
h_1=\dfrac{w}{(1+w)^2}\, ,\\
h_2&=\dfrac{2w^2}{(1+w)^3}\hh w=ye^{-x}\, .
\end{split}
\ee
Then the functions $\Phi_i(y,\lambda)$ have the following integral representations
\be \n{intP}
\Phi_i(y,\lambda)=\int_0^{\infty}e^{-\lambda x} h_i(w) dx\, .
\ee

The functions $h_i(w)$ in the interval $w\ge 0$ are positive and bounded. Their maxima are located at $w=0$ for $i=0$, $w=1$ for $i=1$, and $w=2$ for $i=2$. We denote by $H_i$ these maximal values. Then one has $H_0=1$, $H_1=1/4$, and $H_2=8/27$, respectively.

We always assume that $\lambda$ is real and positive. Then the relation \eqref{intP} shows that for $y>0$, all three functions $\Phi_i(y,\lambda)$ are real and positive, and they are bounded by
\be
0\le \Phi_i(y,\lambda)\le \dfrac{H_i}{\lambda}\, .
\ee


\acknowledgments

This work was supported by the Natural Sciences and Engineering Research Council of Canada. The authors are also grateful to the Killam Trust for its financial support.
One of the authors (V.F.) is grateful to the Yukawa Institute of Theoretical Physics of the Kyoto University for
its support and hospitality while working on this paper. He also thanks Prof. Shinji Mukahyama and the members of the gravity group for interesting and stimulating discussions during his stay as a visiting professor in the Yukawa Institute. The authors thank Ivan Kolar for discussion of the Palais' Principle of Symmetric Criticality, and for sending us his paper on this subject before its publication.

\vskip 2.5cm



%


\end{document}